\documentclass[a4paper,fleqn, usenatbib]{mnras}

\usepackage{multirow}
\usepackage{newtxtext,newtxmath}
\usepackage[dvipsnames]{xcolor}
\usepackage{xcolor}
\usepackage[T1]{fontenc}
\usepackage{ae,aecompl}
\usepackage{aas_macros}

\usepackage{graphicx,graphics}	% Including figure files

\usepackage{amsmath}	% Advanced maths commands

\usepackage{natbib}

\usepackage{scalerel}
\usepackage{tikz}
\usetikzlibrary{svg.path}

\definecolor{orcidlogocol}{HTML}{A6CE39}
\tikzset{
  orcidlogo/.pic={
    \fill[orcidlogocol] svg{M256,128c0,70.7-57.3,128-128,128C57.3,256,0,198.7,0,128C0,57.3,57.3,0,128,0C198.7,0,256,57.3,256,128z};
    \fill[white] svg{M86.3,186.2H70.9V79.1h15.4v48.4V186.2z}
                 svg{M108.9,79.1h41.6c39.6,0,57,28.3,57,53.6c0,27.5-21.5,53.6-56.8,53.6h-41.8V79.1z M124.3,172.4h24.5c34.9,0,42.9-26.5,42.9-39.7c0-21.5-13.7-39.7-43.7-39.7h-23.7V172.4z}
                 svg{M88.7,56.8c0,5.5-4.5,10.1-10.1,10.1c-5.6,0-10.1-4.6-10.1-10.1c0-5.6,4.5-10.1,10.1-10.1C84.2,46.7,88.7,51.3,88.7,56.8z};
  }
}

\newcommand\orcidicon[1]{\href{https://orcid.org/#1}{\mbox{\scalerel*{
\begin{tikzpicture}[yscale=-1,transform shape]
\pic{orcidlogo};
\end{tikzpicture}
}{|}}}}
\newcommand{\be}{\begin{equation}}
\newcommand{\ee}{\end{equation}} 
\newcommand{\bse}{\begin{subequations}}
\newcommand{\ese}{\end{subequations}} 
\newcommand{\bary}{\begin{eqnarray}}
\newcommand{\eary}{\end{eqnarray}}

\newcommand{\Msun}{\rm M_{\sun}}

\newcommand{\lsim}{\mathrel{\hbox{\rlap{\lower.55ex\hbox{$\sim$}} \kern-.3em\raise.4ex\hbox{$<$}}}}
\newcommand{\gsim}{\mathrel{\hbox{\rlap{\lower.55ex\hbox{$\sim$}} \kern-.3em\raise.4ex\hbox{$>$}}}}

\newcommand{\RECAL}{Recal-L025N0752}

\newcommand{\mstar}{ {\rm M_{\star}} }
\newcommand{\dexkpc}{dex~kpc$^{-1}$ }

\newcommand{\Reff}{$R_{\rm eff}$}
\newcommand{\grad}{$\nabla_{{\rm (O/H)}}$}

\newcommand{\eagle}{{\sc eagle}}
\newcommand{\subfind}{{\sc subfind}}
\newcommand{\sigmastar }{$\Sigma_{\star}$}
\newcommand{\oh}{$ 12 + \rm log (O/H)$}

\defcitealias{Tissera2022}{T22}

\title[The metallicity gradients and assembly history]{The metallicity gradients of star-forming regions store information of the assembly history of galaxies}
\author[ Jara-Ferreira et al.]{F. Jara-Ferreira$^{1,2}$\thanks{E-mail: \href{mailto:fujara@uc.cl}{fujara@uc.cl} }, P. B. Tissera$^{1,3}$$^{\orcidicon{0000-0001-5242-2844}}$, E. Sillero$^{1}$, 
Y. Rosas-Guevara$^{4}$$^{\orcidicon{0000-0003-2579-2676}}$,
S. E. Pedrosa$^{5}$,
\newauthor
M. E. De Rossi$^{5,6}$, T. Theuns$^{7}$$^{\orcidicon{0000-0002-3790-9520}}$, L. Bignone$^{5}$
\\\\
$^{1}$Institute of Astronomy, Pontificia Universidad Cat\'olica de Chile, Santiago, Chile.\\ 
$^{2}$N\'ucleo Milenio ERIS, Av. Vicuña Mackenna 4860, 782-0436 Macul, Santiago, Chile.\\
$^{3}$Centro de Astro-Ingenier\'ia, Pontificia Universidad Cat\'olica de Chile, Santiago, Chile.\\
$^{4}$Donostia International Physics Centre (DIPC), Paseo Manuel de Lardizabal 4, 20018 Donostia-San Sebastian, Spain.\\
$^{5}$Instituto de Astronom\'ia y F\'isica del Espacio, CONICET-UBA, Casilla de Correos 67, Suc. 28, 1428, Buenos Aires, Argentina.\\
$^{6}$ Facultad de Ciencias Exactas y Naturales y Ciclo B\'asico Com\'un, UBA, Argentina.\\
$^{7} $Institute for Computational Cosmology, Physics Department, University of Durham , South Road, Durham DH1 3LE, UK.
}

\date{Accepted 2024 March 04. Received 2024 March 04; in original form 2023 September 30}

\pubyear{2024}

\begin{document}
%\label{firstpage}
\maketitle

%academic.oup.com/mnras/article/469/2/2121/3098186
\begin{abstract}

The variations in metallicity and spatial patterns within star-forming regions of galaxies result from diverse physical processes unfolding throughout their evolutionary history, with a particular emphasis in recent events. Analysing MaNGA and \eagle~galaxies, we discovered an additional dependence of the mass-metallicity relation (MZR)  on metallicity gradients (\grad). Two regimes emerged for low and high stellar mass galaxies, distinctly separated at approximately $\mstar >10^{9.75}$.
 Low-mass galaxies with strong positive \grad~appear less enriched than the MZR median, while those with strong negative gradients are consistently more enriched in both simulated and observed samples. Interestingly, low-mass galaxies with strong negative \grad~exhibit high star-forming activity, regardless of stellar surface density or \grad. In contrast, a discrepancy arises for massive galaxies between MaNGA and \eagle~datasets. The latter exhibit a notable anticorrelation between specific star formation rate and stellar surface density, independent of \grad, while MaNGA galaxies show this trend mainly for strong positive \grad. Further investigation indicates that galaxies with strong negative gradients tend to host smaller central black holes in observed datasets, a trend not replicated in simulations. These findings suggest disparities in metallicity recycling and mixing history between observations and simulations, particularly in massive galaxies with varying metallicity gradients. These distinctions could contribute to a more comprehensive  understanding of the underlying physics.
\end{abstract}


\section{Introduction} 

The presence of oxygen radial profiles in the star-forming regions of galaxies is a well-determined observational fact \citep{searle1971, peimbert1975}. The slope of these profile, hereafter metallicity gradients \grad,%\footnote{{\bf{Defined as the gradient of the radial profile of the oxygen abundance, generally used as a proxy of metallicity.}}}
~in  massive, spiral galaxies  tend to become shallower with increasing mass, while a large variety of  \grad~are detected for small galaxies \citep[][]{zaritsky1994, Ho2015}. %This mass dependence is reduced when using the metallicity gradients normalised by a characteristic scale such as the half-mass radius \citep{diaz1989}. This metallicity gradient will be called instrisic \grad, hearafter. 
More recently, large surveys of galaxies using integral field spectroscopy (IFS) such as CALIFA \citep{sanchez2013Califa}, MaNGA \citep{Bundy2015} and SAMI \citep{poet2018} have allowed a more detailed characterization of the metallicity  distribution in  nearby galaxies. From this data, a weak anti-correlation between \grad~and stellar mass  for the low-mass end has been detected, so that  that slopes  became shallower or even positive\footnote{For these galaxies, the central gas-phase metallicity is shallower than in the outskirts. These \grad~will be also referred as inverted metallicity gradients.}  \citep{sanchez2013Califa, sanchezmenguiano2016, belfiore2017}. Positive gradients have also been observed at high redshift, particularly, in connection with gas-rich galaxies that are frequently undergoing interactions \citep[e.g][]{troncoso2014,  jones2015, carton2018,wang2019b,wang2022a}.
%On the other hand, galaxies in interactions have been reported to have diluted central regions and to often exhibit weaker negative or even positive  metallicity  gradients  \citep[e.g.][]{kewley2006, rupke2010}. %Using MUSE observations, \citet{SanchezMen2018} 

Metallicity gradients are expected to be shaped by the action of different physical processes \citep{tinsley1980,freeman}, such as star formation,  energy feedback as well as gas inflows, mergers, interaction and probably environmental mechanisms such as ram pressure and tidal striping \citep{bahe2017}. Hence, they encode information on these processes, which can provide clues for understanding the history of evolution of galaxies \citep{pilkington2012,tissera2016a,maiolino2019,sharda2023}.

Within the standard inside-out disc formation scenario \citep{fall1980}, negative metallicity gradients are  expected to form as the star formation takes place in high-density gas clouds, starting from the central regions and moving out progressively to the outskirts of the discs.  If the gas accretion continues in a steady fashion, this gas will settle onto the disc, fuelling the ongoing star formation activity and setting off supernovae (SNe) feedback.
Energy feedback contributes to modify the thermodynamical properties of the interstellar medium (ISM) \citep[e.g.][]{wiersma2009,scan08}, to mix chemical elements or even to eject part of them  via galactic winds \citep[e.g.][]{stin2006,hopkins2013}. Previous works have shown that SNe feedback is one of the main processes that regulate the slope of metallicity gradients together with the availability of gas to form stars and the efficiency of transformation of gas into stars \citep{pilkington2012,lilly2013,stinson2013,gibson2013,ma2017,molla2017, belfiore2019,tissera2019, wang2022}. 

 However, galaxies are rarely isolated entities, but have ongoing interactions with their surroundings. These interactions can lead to the capture of gas and stars from neighbouring galaxies as well as the stripping of material from the galaxies themselves, among other possible effects \citep[e.g.][]{karademir2019,pallero2022,rodriguez2022}. These mechanisms could affect the metallicity distribution in the star-forming regions, shaping the metallicity gradients and the level of enrichment of galaxies \citep{franchetto2021,sharda2021,maier2022}.
Previous works have shown that galaxy-galaxy interactions and mergers can generate tidal torques that affect the distribution of gas and stars, and hence, the star formation activity and metallicity distribution across galaxies \citep{rupke2010, dimatteo2013,sillero2017,moreno2019}. These tidal torques are also expected to trigger bars, which have been  shown to be efficient at transporting gas into the central regions as the gas loses angular momentum \citep{weinberg1985,athanassoula2003}. This gas has the potential to boost star formation activity in these regions and hence, to modify the chemical abundances of the ISM as the galaxies draw closer to each other \citep{katz1992,tissera2000,lambas2003,torrey2012,moreno2019}.
%As a consequence, these interactions could also lead to double starbursts, one induced during the first pericentre and the second one, during the merger itself depending on the characteristics of the interaction \citep{tissera2000}. 

These gas inflows induced during galaxy-galaxy interactions can also explained the dilution of the central oxygen abundance as shown by numerous observational and numerical works \citep[e.g.][]{kewley2006,rupke2010, kewley2010,perez2011,tissera2016a,bustamante2018,moreno2019,bustamante2020}. Depending on the characteristics of the interactions and gas richness of the encountering galaxies, strong starbursts could be generated, which on their turn, can induce metal-loaded outflows \citep[e.g][]{scan09,stinson2013}. This sequence of events is expected to change the chemical distribution of the ISM. In the first stage, the transport of material from outer regions to the centre will dilute the metallicity but as new stars are formed, the $\alpha$-enrichment will increase by the contribution of Type II SNe (SNeII). Depending on the strength of the starbursts and the potential well of the galaxies, metal-loaded galactic winds can potentially diminish the level of enrichment and modify the thermodynamical properties of the ISM by expelling enriched material or by redistributing it via galactic fountain, regulating the subsequent star formation activity.  \citet{sillero2017} showed that if a gas inflow is triggered in a short time-scale,  $\sim 0.5$ Gyr, then, it could both induce a starbursts and change metallicity slopes. Active Galactic Nuclei (AGN) activity could be also triggered during these events and have been claimed to be very efficient at quenching star formation in massive galaxies \citep[e.g.][]{rosasguevara2019}. However, it is not yet clear enough how AGN feedback can modify the thermodynamical properties of the ISM and its chemical characteristics across the discs \citep{hopkins2013,rosas-guevara2015}. 

Collectively, the aforementioned physical processes govern the star formation activity, the degree of enrichment, and the distribution of chemical elements within galaxies of varying stellar masses across different evolutionary phases. Consequently, these processes can potentially imprint discernible effects on relations such as the mass-metallicity relation, MZR.

Although, the MZR has been extensively studied \citep[][and references there in]{maiolino2019}, it remains to be fully understood which physical mechanisms could drive it and what is the origin of its scatter as a function of time \citep{curti2023}. \citet{ellison2008} found that part of the scatter could be related to the size and the star formation activity so that galaxies with larger star formation rates, SFR, tend to have lower metallicity \citep[see also][]{mannucci2010,laralopez2010}. A similar correlation of the scatter with the HI mass has been also reported \citep{bothewell2013}.

Observations have also found  that galaxy interactions could shift the location of galaxies on the MZR. \citet{micheldansac2008} studied pairs of galaxies selected from the SDSS-DR4 and found that merging galaxies exhibited an excess of metallicity in massive galaxies, while smaller galaxies showed a dilution of metallicity. Similarly, \citet{omori2022} reported a dilution of metallicity in very close pairs compared to systems with members located at larger distances. Both studies attributed the variations in chemical enrichment in pairs to the effects of the interaction itself, as it is actually predicted by  numerical simulations \citep{perez2006, perez2011,torrey2012,tissera2016,sillero2017}. Recent observational studies of the resolved MZR have reported that a key property influencing its scatter is the stellar surface density or stellar compactness, \sigmastar~\citep{boardman2022}. Furthermore, these studies indicate that metallicity gradients may not simply be a by-product of the resolved properties, which implies a direct increase of the chemical abundances for increasing \sigmastar,  but could have an intrinsic origin related to the action of other processes such as gas inflows and mergers \citep{baker2023,boardman2023}.

From a numerical point of view, \citet{torrey2019} found that the scatter in the simulated MZR estimated for galaxies in Illustris TNG50 is correlated with the gas fraction and star formation activity.
The \eagle~simulations have been extensively used to study the MZR and its scatter \citep[e.g.][]{crain2015,lagos2016,derossi2017,trayford2019b}. In particular \citet{vanloon2021} found that the scatter of the MZR in \eagle~simulations was related to variations of long time-scales, probably reflecting the combined action of inflows, outflows and star formation. These results agree with previous findings from analytical modelling of the relation between SFR, metallicity and gas fraction and the time variation \citep[e.g.][]{forbes2014, wanglilly2021}.
Recently,  \citet{zenocratti2020,zenocratti2022} studied the MZR of \eagle~galaxies  with stellar mass in the range $[10^{9},10^{10}] \Msun$. These authors studied the level of enrichment in relation to galaxy morphology, finding that, at  given stellar mass, more disc-dominated galaxies tended to have lower level of enrichment than  dispersion-dominated systems. They related this trend to the late assembly history of disc-dominated galaxies and hence, the latter accretion of low-metallicity gas, compared to spheroidal-systems.

Therefore, the simulated and observational results on the MZR and the the mass- metallicity gradients  relation, MZGR, have prompted us to further analyse the information contained in the \grad~of star-forming regions in relation to their level of enrichment. In this paper, we explore the potential link between the  MZR and  MZGR to unravel fossil records of their evolution,  possible stored by the metallicity gradients, by exploring a possible secondary dependence of the MZR on \grad.

To achieve this goal, we resort to galaxies from the MaNGA survey \citep{sanchez2013Califa} and simulated galaxies from the higher resolution run, \RECAL, of the \eagle~project \citep{schaye2015}. We conduct a comparative analysis between galaxies with different  \grad, grouping them according to their metallicity gradient and stellar mass. For this purpose, we follow  \citet[][hereafter T22]{tissera2022} who studied the metallicity gradient evolution and the relation with the merger and gas inflow histories. 
These authors found that galaxies with recent strong accretion (i.e. either mergers or gas inflows) tend to be more star-forming and have strong negative or positive gradients and that after about 2 Gyr they recovered their weak metallicity gradients. 
This work is be taken as reference and, thus we adopt their definitions as stated throughout the paper. 
%as strong negative ($< -0.02~ $\dexkpc), strong positive ($ >0.02 $~\dexkpc) and weak metallicity gradients, in both the MaNGA and \eagle~samples. 
%adopting  for consistency with \citet{tissera2022} since use their data catalogue.

This paper is organised as follows. In Section \ref{sec:data} we summarises the main characteristics of selected galaxies from  the MaNGA survey and the  \eagle~simulation. In Section \ref{sec:analysis}, we analyse the relations between the metallicity gradients, specific star formation rate, sSFR, and strong accretion history and AGN activity. Section \ref{sec:conclusions} summarises our main findings.

\section{Simulated and observed data} \label{sec:data}
\subsection{ The MaNGA galaxies} \label{sec:manga}

We use the galaxy catalogue built with Pipe3D \citep{sanchez2016} on the Data Release 16 of the Mapping Nearby Galaxies at Apache Point Observatory \citep[MaNGA,][]{Bundy2015}. The MaNGA survey is  one of the projects included in the Sloan Digital Sky Survey  IV  \citep[SDSS-IV:][]{Blanton2017}. This survey uses integral field units (IFU) and has generated a database of about 10000 spatially resolved nearby galaxies, with a wide variety of morphologies and a relatively uniform mass distribution. Only galaxies with inclinations lower than 75 degrees have been considered for the analysis.
%XXXXXVOLVER A poner lo de la morfologia --mirar la tabla 2
From the catalogue, we also obtained the stellar mass, $\mstar$, the star formation rate, SFR, and the effective radius, \Reff\footnote{A detailed description of the estimation of the observed gradients and parameters can be also found in \citet{Sanchez2022}}. 
%The sample covers a range of redshifts between $z = 0.01$ and $z = 0.15$ \citep{wake2017}. 
%The Pipe3D catalogue  described in \citet{sanchez2016} provides spatially resolved measurements of the main galactic characteristics of its members, such as  stellar mass, chemical abundances, stellar formation activity, kynematics, among others, for about 2000 galaxies. Such data is available in the SDSS Science Archive Server\footnote{SAS: \url{https://data.sdss.org/sas/}} and can also be download from the SkyServer DR16 site\footnote{\url{https://skyserver.sdss.org/dr16/en/home.aspx}}. We visually inspected the data and discarded stronger metallicity gradients since most of them were detected in very distorted galaxies or the fittings were not accurate enough (i.e. low S/N).

We  selected galaxies based on four criteria: (i) redshift $z < 0.05$, (ii) stellar masses, $\mstar \in [10^9, 10^{11}] \Msun$, (iii) metallicity gradients between $-0.2$ and $0.2$ dex \Reff$^{-1}$, and (iv) sSFR within the range covered by the \eagle~galaxies  for each of the three defined mass intervals (see Section \ref{sec:eagle}). This fourth selection criterion constrains the  sSFR to $[10^{-10.44}, 10^{-9.40}]$yr$^{-1}$ for low-mass, $[10^{-11.32}, 10^{-9.57}]$yr$^{-1}$ for intermediate mass and $[10^{-11.46}, 10^{-9.75}]$yr$^{-1}$ for high mass galaxies. By applying these criteria, we obtained our main sample of 1765 galaxies, hereafter referred to as the MaNGA sample. 

We selected the metallicity gradients measured in the radial range [0.5, 1.5]\Reff, using the abundance of oxygen as a tracer. Our reference metallicities were measured using the O3N2 calibrator \citep{marino2013}, which is a direct method that depends on two strong emission lines; [NII] $\lambda6583$ and [OIII] $\lambda5007$. However, we also explored other calibrators finding that most of them produced similar relations, as shown in Appendix~\ref{metallicitycalibrators}.

%---------------

We acknowledge the fact that all the metallicity gradients used in this paper were obtained by fitting a single linear regression to the metallicity profiles within the radial range $[0.5-1.5]$\Reff. Recent observational results suggest that more than one linear fit might be required to fully describe the metallicity profiles \citep[e.g.][]{diaz1989,sanchezmenguiano2016,barreraballesteros2022}. Numerical simulations also report the existence of  breaks from a single power law, which could be reflecting the action of different physical parameters such as accretion, gas inflows, galactic fountains \citep[e.g.][Tapia in preparation]{garcia2023}.  Because in this study, we specifically concentrate on the regions dominated by the disc components, we anticipate  no significant changes because of this, as breaks are typically identified in the inner regions or outskirts \citep{sanchezmenguiano2016,tapia2022}. 
  
Additionally, we estimated the mass of central black hole, M$_{\rm{BH}}$, by using the known correlation between  M$_{\rm{BH}}$ and the stellar velocity dispersion of the central region, $\sigma_c$. The stellar velocity dispersion in the central $2.5$ arcsec/aperture  was used as given by Pipe3D. The correlation reported by \citet[][]{tremaine2002} was followed to estimate M$_{\rm{BH}}$. The authors proposed empirical values for the coefficients of a linear correlation with the form $\log(\rm{M}_{\rm{BH}}) = \alpha + \beta\log(\sigma/\sigma_0)$, which are $\alpha = 8.13 \pm 0.06$ and $\beta = 4.02 \pm 0.32$, for $\sigma_0 = 200$ km s$^{-1}$. 
Hence, we obtained M$_{\rm{BH}}$ with a median error of  $\sim \log(0.12)$.
%We spread the errors of the $\sigma_c$ measurements from Pipe3D and the fit from T00, obtaining a median error of $\sim \log(0.12)$. %This is lower than any error bar associated with median trends in figures involving M$_{\rm{bh}}$ of MaNGA galaxies. 
%To ensure robustness in our analysis, we also estimated the M$_{\rm{bh}}$ of MaNGA galaxies under the correlations proposed by \citet{mcconnell2013}  and \citet{saglia2016}. 
%These authors proposed a correlation which depends on galaxy.  To apply it to our sample we defined Eearly TGs as galaxies with $v_{\rm{rot}}/\sigma<0.4$ and LTGs those with $v_{\rm{rot}}/\sigma\geq 0.4$. Our results remain unchanged with this different fits for the M$_{\rm{bh}}-\sigma$ correlation.}}

\subsection{ The \eagle~galaxies }
 \label{sec:eagle}

We use a set of galaxies selected from  the \eagle~Project\footnote{The database is publicly available, see \citet{mcAlpine2016}.}, which is a suite of  cosmological hydrodynamical simulations \citep{schaye2015,schaller2015}.
\eagle~simulations are in agreement with  a Lambda Cold Dark Matter ($\Lambda$CDM) universe with the cosmological parameters\footnote{ We have re-scaled the MaNGA parameters  \citep{sanchez2016} to match the cosmology adopted by the \eagle~project.} of the \citet{planck2014} namely $\rm \Omega_{\Lambda} = 0.693$, $\rm \Omega_{m} = 0.307$, $\rm \Omega_{b} = 0.04825$, $h = 0.6777$ ($H_{\rm 0} = 100\ h$ $\rm km\ s^{-1}\ Mpc^{-1}$), $\rm \sigma_{8} = 0.8288$, $n_{\rm s} = 0.9611$, and $Y = 0.248$. 

In this work we use the \RECAL~run,  which corresponds to a volume of  $25$ cMpc (co-moving mega-parsecs) size length, and uses $752^{\rm 3}$ initial baryonic and dark matter particles.
%The  gravitational softening ($0.35$ pkpc, proper kilo-parsecs) is kept constant in proper units below $z=2.8$; at higher $z$ the softening is kept constant in co-moving units
%at 1.33~ckpc. 
The mass resolution is $2.26 \times 10^5~\Msun$ and $1.21 \times 10^{6}~\Msun$ for the initial gas and dark matter particles, respectively \citep[see][for more details]{schaye2015}. This corresponds to the highest resolution within the \eagle~suite.  And, in particular, the recalibrated model predicts a MZR slope in better agreement with observations of low-mass galaxies \citep{schaye2015, derossi2017}.

%\subsubsection{Subgrid Physics}
The simulations were performed with the so-called {\sc ANARCHY} version of {\sc p-gadget-3} \citep[][see for more details on the subgrid physics]{schaye2015,crain2015}. %Details of the subgrid  physics models which are relevant to this work can be found in \citet[][and references therein]{tissera2019}.
Briefly, it includes a star formation algorithm  which forms stars  stochastically, following the model
of \citet{schaye2008},  from cold and dense gas with a metallicity dependent
star formation density threshold \citep{schaye2010}. The cooling and photo-heating models are implemented following \citet{wiersma2009a}. 
 The stellar feedback is treated stochastically, adopting the thermal
injection scheme described in \citet{dallavecchia_schaye2012}. The chemical model follows eleven chemical elements \citep{wiersma2009}. The \eagle~simulations adopt the initial mass function  of \citet{chabrier2003}, with  minimum and  maximum mass cut-offs of  $0.1~\Msun$  and $100~\Msun$, respectively.

The AGN feedback model locates black holes, BHs, with an initial mass of $\rm m_{\rm seed} = 1.48 \times 10^5~\Msun$  at the centres of haloes with a dynamical mass exceeding $ \sim 10^{10}~\Msun$ \citep{springel2005}. The growth of BHs occurs through gas accretion and mergers. The accretion rates of the supermassive black holes, SMBH, are computed using the modified Bondi-Hoyle accretion rate formula as explained by  \citet{rosas-guevara2015, schaye2015}. %To adjust the Bondi-Hoyle accretion rate in high circulation flows, a viscosity parameter is introduced. 
%Additionally, the accretion rates are constrained by the Eddington rate. 
The AGN feedback parameters were calibrated to replicate the stellar mass galaxy function, the $\rm \mstar-M_{\rm BH}$ and the scaling relation between stellar mass and the mass of the central black hole for galaxies observed in the Local Universe as described in \citet{crain2015}.
%In \RECAL~the change in temperature of the surrounding ISM is modified by  $\Delta T = 10^9 \rm~K$.

%The BH accretion regime, which is useful to classify the activity state of the galaxy, can be quantified through the Eddington ratio, $\lambda_{\rm{edd}}$,
%The Eddington rate can be estimated as 
%\begin{equation}
%    \lambda_{\rm{edd}} = \frac{\epsilon\cdot  c \cdot \dot{M}_{\rm{BH}} \cdot \sigma_{\rm{T}}}{4\pi \cdot G\cdot M_{\rm{BH}} \cdot m_p}, 
%\end{equation}
%where $\epsilon$ is the efficiency (fixed to 0.1), $\sigma_{\rm{T}}$ is the Thompson cross-section and $m_p$ is the mass of the proton. $\dot{M}_{\rm{BH}}$ and $M_{\rm{BH}}$ are the SMBH mass accretion rate and mass, respectively. 
To explore the possible impact of AGN feedback we use the Eddington rates, $\lambda_{\rm{Edd}}$\footnote{The Eddington rate can be estimated as 
\begin{equation}
    \lambda_{\rm{edd}} = \frac{\epsilon\cdot  c \cdot \dot{M}_{\rm{BH}} \cdot \sigma_{\rm{T}}}{4\pi \cdot G\cdot M_{\rm{BH}} \cdot m_p}, 
\end{equation}
where $\epsilon$ is the efficiency (fixed to 0.1), $\sigma_{\rm{T}}$ is the Thompson cross-section and $m_p$ is the mass of the proton.}, and $\rm M_{\rm BH}$. In \eagle~the  BH accretion regimes are controlled by  the $\lambda_{\rm{Edd}}$  as described by \citet{Rosas-Guevara2016}, where galaxies with $\lambda_{\rm{Edd}} > 10^{-2}$ are assumed to be highly accretive, X-rays luminous sources; galaxies with $10^{-4}< \lambda_{\rm{Edd}} < 10^{-2}$ are expected to be radiatively inefficient % and in  the regime of Advection Dominated Accretion Flows 
\citep[][]{narayan1994, abramowicz1995}; and galaxies with $ \lambda_{\rm{Edd}} < 10^{-4}$ are considered as inactive. Nevertheless, we stress the fact that $\lambda_{\rm{Edd}}$ show large variability as discussed by  \citet{rosasguevara2019}.
Thus, whilst still showing the trend with $\lambda_{\rm{Edd}}$, we focus the main analysis of the AGN impact on the sSFR and metallicity gradients by using the $\rm M_{\rm BH}$ provided by the \eagle~database. The detailed estimations of $\rm M_{\rm BH}$ can be found in \citet{Rosas-Guevara2016}.

\subsubsection{\eagle~ galaxy sample}
\label{sec:eaglesample}
Galaxies were identified by using the Friends-of-Friends
\citep[FoF,][]{davis1985} and \subfind~\citep{springel2001a, dolag2009} algorithms. Our analysis is restricted  to central galaxies. We adopted the galaxy catalogue  built   by  T22. 
%The catalogue include galaxies identified up to $ z \sim 2$
%\subsubsection{Main properties of the simulated galaxies} \label{sec:sim_glx_z}
These authors identified the discs and  spheroidal components based on the angular momentum and the binding energy (AM-E method) of the stars and gas as described by \citet{tissera2012}. 
%The bulge, disc and stellar haloes are separated according to  $\epsilon= J_{\rm z}(E)/J_{\rm z,max} (E)$. Here, $J_{\rm z}(E)$ is the angular momentum along the $z$-axis of the particle with binding energy $E$, and $J_{\rm z,max} (E)$ is the maximum value of $J_{\rm z}$ over all particles with the same binding energy. 
This decomposition also allows the estimation of the stellar disc-to-total mass, D/T, which will be used as a morphology estimator \citep[e.g.][]{pedrosa2015,correa17}.

For each simulated central galaxy in the \eagle~sample, the stellar mass ($\mstar$), the stellar half-mass radius (\Reff), the sSFR, the gas fraction ($f_{\rm gas}=\rm M_{ \rm gas}/(\mstar + \rm M_{ \rm gas})$) and the stellar surface density (\sigmastar) were computed.
Additionally, we adopted the parameter
$\tau_{25}$ defined by T22, which measures the time elapsed since the last major increase of stellar mass, by more than 25 per cent, between two available snapshots. Hence, we are agnostic with respect to the processes that could caused it (e.g. gas inflows or mergers). These parameters have been estimated by using the gas or stellar particles within 1.5 times the optical radius\footnote{These parameters are taken from T22, where the optical radius is defined as the one that encloses $\sim80$ per cent of stellar mass contained with a fixed aperture of 30 kpc.}.
 
The \grad~of the star-forming regions in the discs components are calculated by using the the oxygen abundances of the gas particles that satisfy all simulation requirements for being converted to stars but have not being yet transformed. As explained by T22, galaxies with stellar masses below $10^9 M_{\sun}$, hence, having typically fewer than 1000 gas particles, are not considered in order to mitigate numerical effects on the estimation of the metallicity gradients due to limited numerical resolution.
    
To enhance the realism of the comparison with observations, the oxygen profiles are weighted by the SFR of the corresponding regions. Then, the logarithm azimuthal-averaged radial profiles are fitted by a linear regression within the radial range $\in [0.5, 1.5]$\Reff~in those galaxies with more than 100 star-forming gas particles.
Galaxies with $R_{\rm eff} < 1$~kpc (about three times the gravitational softening) and  stellar masses lower than $10^9~\Msun$ were excluded from the sample to mitigate the effects of low numerical resolution on the determination of the metallicity profiles.

The global metallicity of the \eagle~sample is estimated as the global oxygen abundance defined by  the star forming regions in each galaxy \citep{degraaff2022} within 1.5~\Reff. \citet{derossi2017} reported that galaxies in  the \RECAL~run have a MZR in good agreement with observations. We note that our estimations are all referred to 1.5~\Reff, instead of adopting a fixed aperture radius \citep{zenocratti2022}. Also, our sample is restricted to star-forming galaxies, which satisfy the minimum number of star-forming gas particles to reliably estimate the  metallicity gradients, so galaxies with low star-formation activity will not be included. %These are two main differences with respect to the work of \citet{zenocratti2022} that should be bear in mind. 

Throughout our analysis, we use results from  \citet{tissera2022}. To ensure consistency, we adopt the same criteria for grouping galaxies into stellar mass and metallicity gradient intervals. Three mass intervals are defined: $[10^9, 10^{9.75})\Msun$, [$10^{9.75}, 10^{10.25}]\Msun$ and $(10^{10.25}, 10^{11}]\Msun$.  Additionally, galaxies are analysed according to  three \grad~subsamples: strong negative ($< -0.02$~\dexkpc), strong positive ($ >0.02 $~\dexkpc) and weak metallicity gradients.  This procedure is applied to both simulated and observational data. While we use metallicity gradients in \dexkpc to define the samples for direct comparison with the results of \citet{tissera2022}, we also estimated them normalised by \Reff~and checked that this does not affect the trends. In Table~\ref{table2}, the median properties of galaxies in each mass interval are summarised.

We note that, in \eagle~we only consider central galaxies (i.e. the most massive galaxy within a dark matter halo). However, for MaNGA we have not distinguished between central and satellite galaxies. These could introduce same differences considering that satellite galaxies seem to be slightly more enriched and less star-forming, particularly in high density regions \citep{pasquali2012}. However, since we are focusing on star-forming galaxies in both observations and simulations, we expect the comparison not to be strongly affected by the possible presence of satellite galaxies in the MaNGA sample.

\begin{figure}
\resizebox{8cm}%{!}{\includegraphics{Figuras/Fig1.pdf}}
{!}{\includegraphics{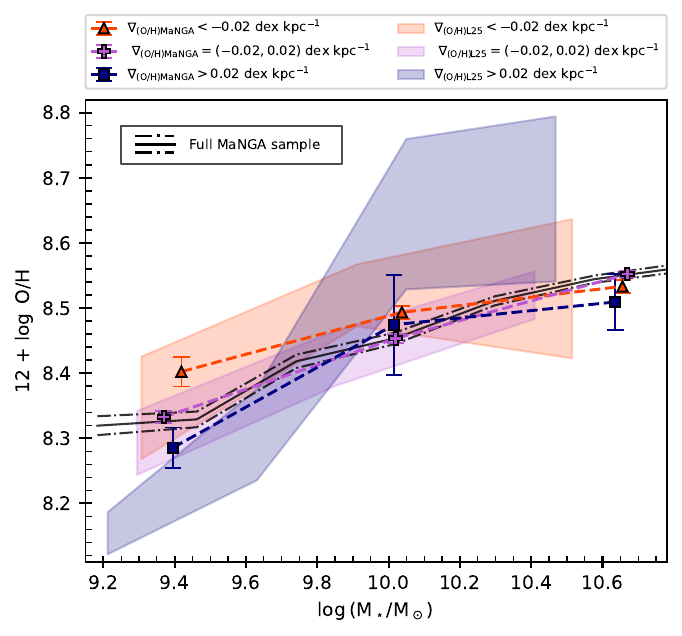}}
\caption{The mass-metallicity relation as a function of the metallicity gradients of galaxies in our MaNGA and \eagle~samples.  The median MZR for galaxies with strong negative (red), weak (pink) and strong positive (blue) metallicity gradients for MaNGA (symbols and lines) and the \eagle~(shaded areas) samples are denoted according to information in the inset. The median MZR for the full MaNGA sample (black, solid line) and the $95\%$ confidence intervals (black, dotted-dashed lines). Error bars and the widths of the shaded areas were estimated by using a bootstrap technique.}
\label{mzrgrad}
\end{figure}

\begin{table*}
\caption{Summary of the main properties in our \eagle~subsamples at $z=0.0$. From the left to the right columns the fraction with respect to the complete sample, the medians of the gas-phase metallicity gradient, global metallicity, effective radius, specific star formation rate, gas fraction and Eddington ratio are displayed. The medias are shown for low-mass galaxies, $\mstar < 10^{9.75}~\Msun$, intermediate mass galaxies, $\mstar = [10^{9.75}, 10^{10.25}]~\Msun$,  and high-mass galaxies, $\mstar > 10^{10.25}~\Msun$. The strong negative group (S-Neg.) corresponds to galaxies with $\nabla_{\rm{O/H}}<-0.02$ dex~kpc$^{-1}$, the gradients within the range of $(-0.02,0.02)$ dex kpc$^{-1}$ are defined as weak group, and the strong positive group (S-Pos.) corresponds to galaxies with $\nabla_{\rm{O/H}}>-0.02$ dex~kpc$^{-1}$ \citep{tissera2022}. The associated uncertainties correspond to 95 per cent confidence limits calculated by using a bootstrap technique.}
\label{table2}
\resizebox{\textwidth}{!}{
%\begin{tabular}{llcllllll}
\begin{tabular}{llccccccc}
\multicolumn{2}{c}{\multirow{2}{*}{}} & \multicolumn{7}{c}{\eagle~sample $z= 0.0$}             \\ \cline{3-9} 
\multicolumn{2}{c}{}                                     &$f_\%$ &\grad  & \oh & \Reff (kpc) & $\log$ (sSFR) & $f_{\rm{gas}}$ & $\log(\lambda_{\rm{Edd}})$ \\ 
\textbf{}
\multirow{3}{*}{\begin{tabular}[c]{@{}l@{}}Low \\ Mass\end{tabular}}     & S-Neg.& $0.130$      & $-0.039\pm 0.008$     &       $8.33 \pm 0.08$     &   $3.54\pm 0.67$  &    $-9.87\pm 0.13$  & $0.35\pm 0.08$   & $-6.44\pm 1.00$              \\
                                & Weak&  $0.326$                 &   $-0.000\pm 0.004$    & $8.33 \pm 0.05$            &   $4.25\pm 0.57$   &   $-9.91\pm 0.07$    &  $0.41\pm 0.05$    &    $-7.14\pm 0.50$       \\
                                & S-Pos.& $0.043$       &      $0.050 \pm 0.010$&    $8.17 \pm 0.05$          &   $3.28\pm 0.60$    & $-9.95\pm0.25$   &  $0.34\pm 0.08$    &    $-9.14\pm 2.45$           \\ \hline
\multirow{3}{*}{\begin{tabular}[c]{@{}l@{}} Inter. \\ Mass\end{tabular}}          & S-Neg.&$0.043$        &   $-0.030 \pm 0.010$   &       $8.47\pm 0.04$      &   $3.65\pm 0.90$   &   $-10.60\pm 0.32$   &   $0.11\pm 0.06$   &       $-5.48 \pm 0.97$        \\
                                & Weak & $0.239$                 & $0.003\pm 0.003$     &   $8.43\pm 0.03$          &   $4.20\pm 0.64$   &    $-10.14\pm 0.12$  &   $0.22\pm0.04$   &   $-5.05\pm 0.76$              \\
                                & S-Pos.&$0.033$        &  $0.030\pm 0.010$    &     $8.53\pm 0.16$         &   $3.20\pm 1.02$    &    $-10.42\pm 0.14$   &  $0.14\pm 0.06$    &     $-6.26 \pm 1.53$           \\ \hline
\multirow{3}{*}{\begin{tabular}[c]{@{}l@{}}High \\ Mass\end{tabular}}           & S-Neg.& $0.022$       &  $-0.034\pm 0.001$    &     $8.66\pm 0.08$       &  $3.52\pm 0.19$   &  $-10.23\pm 0.03$   &   $0.06\pm 0.02$   &     $-4.07\pm 0.62$         \\
                                & Weak & $0.130$                 &   $0.0004\pm 0.0004$   &    $8.52\pm 0.03$          &   $5.21\pm 1.05$  &  $-10.44\pm 0.23$    &  $0.13\pm 0.04$    &       $-4.05\pm0.78$         \\
                                & S-Pos. & $0.033$      &  $0.029\pm 0.001$    &    $8.60\pm 0.11$         &    $2.87\pm 0.99$  &   $-10.77\pm 0.41$    &  $0.08\pm 0.05$   &     $-4.43\pm 1.66$            \\
                                \hline
\end{tabular}
}
\end{table*}

\begin{table*}
\caption{Summary of the main properties in our MaNGA and \eagle~subsamples. From the left to the right columns the fraction with respect to the complete sample, the medians of the gas-phase metallicity gradient, the global metallicity, the effective radius (kpc), and the specific star formation rate (yr$^{-1}$)are displayed. Additionally, $v_{\rm rot}/\sigma$ and D/T have been included for the MaNGA and \eagle~samples, respectively. The medias are shown for  low-mass galaxies, $\mstar < 10^{9.75}~\Msun$, intermediate mass galaxies, $\mstar = [10^{9.75}, 10^{10.25}]~\Msun$,  and high-mass galaxies, $\mstar > 10^{10.25}~\Msun$. The strong negative group (S-Neg.) corresponds to galaxies with $\nabla_{\rm{O/H}}<-0.02$ dex~kpc$^{-1}$, the gradients within the range of $(-0.02,0.02)$ dex kpc$^{-1}$ are defined as weak group, and the strong positive group (S-Pos.) corresponds to galaxies with $\nabla_{\rm{O/H}}>-0.02$ dex~kpc$^{-1}$ \citep{tissera2022}. The associated uncertainties correspond to 95 per cent confidence limits calculated by using a bootstrap technique. }
\label{table1}
\resizebox{\textwidth}{!}{%
%\begin{tabular}{llllllllllll}
\begin{tabular}{llcccccccccccccc}
\multicolumn{2}{c}{\multirow{2}{*}{}} & \multicolumn{7}{c}{MaNGA sample}        & \multicolumn{7}{c}{\eagle~sample $z= [0.0,0.18]$}        \\ \cline{3-16} 
\multicolumn{2}{c}{}                                     & $f_\%$&\grad  & \oh & \Reff  & $\log$ (sSFR)& $\log($M$_{\rm{BH}})$& v$_{\rm{rot}}/\sigma$&$f_\%$& \grad  & \oh & \Reff  & $\log$ (sSFR) & $\log($M$_{\rm{BH}})$ & D/T \\ \hline
\multirow{3}{*}{\begin{tabular}[c]{@{}l@{}}Low \\ Mass\end{tabular}}     & S-Neg.&   $0.035$    & $-0.031\pm 0.003$     &       $8.41 \pm 0.02$     &   $2.46\pm0.26$  &    $-10.17\pm 0.04$& $5.61 \pm 0.38$& $0.46\pm0.04$&$0.144$ &   $-0.06\pm 0.01$   & $8.20\pm 0.10$      &   $4.52\pm 0.70$   &    $-9.90\pm 0.06$ & $5.56\pm0.07$&$0.21\pm0.03$\\
                                & Weak &     $0.167$             &   $-0.002\pm 0.001$    & $8.34 \pm 0.01$            &   $3.29\pm 0.16$   &   $-10.12\pm 0.02$    &
                                $5.42\pm 0.14$
                                & $0.51\pm0.02$&
                                $0.252$&  $0.000\pm 0.003$    &    $7.85\pm 0.05$   &   $6.10\pm 0.50$   &  $-9.85\pm 0.04$  & $5.68\pm0.07$ & $0.30\pm0.03$ \\
                                & S-Pos. &   $0.019$    &      $0.030 \pm 0.003$&    $8.31 \pm 0.03$          &   $2.50\pm 0.47$    & $-10.08\pm0.07$ & $6.53\pm 0.44$ & 
                                $0.48\pm0.05$
                                &$0.079$ &  $0.05\pm 0.01$    &    $7.48\pm 0.16$         &  $4.55\pm 0.61$   &   $-9.90 \pm 0.09$ & $5.62\pm 0.17$
                                &$0.20\pm0.04$
                            \\ \hline
\multirow{3}{*}{\begin{tabular}[c]{@{}l@{}} Inter. \\ Mass\end{tabular}}          & S-Neg.&   $0.098$     &   $-0.030 \pm 0.001$   &       $8.47\pm 0.01$      &   $3.45\pm 0.20$   &   $-10.23\pm 0.04$&     $5.51\pm0.19$&$0.51\pm0.02$ &$0.050$   &   $-0.036\pm 0.010$   &       $8.22 \pm 0.17$      &    $3.86\pm 0.89$  &  $-10.28\pm 0.20$ & $6.34\pm0.25$ &$0.39\pm0.05$  \\
                                & Weak &       $0.282$           & $-0.006\pm 0.001$     &   $8.44\pm 0.01$          &   $3.88\pm 0.15$   &    $-10.26\pm 0.03$& $5.70\pm0.11$
                                
                                &$0.50\pm0.01$&$0.241$  &   $0.002\pm0.002$   &   $7.91\pm 0.05$          &  $5.34\pm 0.63$    &  $-10.12\pm 0.07$ & $6.45\pm0.12$&$0.38\pm0.03$   \\
                                & S-Pos. &   $0.003$    &  $0.030\pm 0.008$    &     $8.42\pm 0.07$         &   $1.68\pm 0.30$    &    $-10.60\pm 0.34$  &
                                $5.83\pm0.54$ &
                                $0.46\pm0.15$
                                
                                &$0.047$ &  $0.040\pm 0.007$    &     $7.63 \pm 0.15$        &  $3.88\pm 0.69$    &   $-10.19\pm 0.09$ & $6.08\pm0.21$ &$0.37\pm0.05$  \\ \hline
\multirow{3}{*}{\begin{tabular}[c]{@{}l@{}}High \\ Mass\end{tabular}}           & S-Neg. &  $0.069$     &  $-0.027\pm 0.001$    &     $8.52\pm 0.01$       &  $4.76\pm 0.30$   &  $-10.24\pm 0.06$& $5.94\pm0.20$&$0.49\pm0.02$ &$0.032$   &   $-0.031\pm 0.004$   &     $8.13\pm 0.12$       &  $3.26\pm 1.23$    &    $-10.60\pm 0.20$ & $7.58\pm0.14$ &$0.26\pm0.08$\\
                                & Weak  &    $0.323$             &   $-0.006\pm 0.001$   &    $8.53\pm 0.01$          &   $5.19\pm 0.16$  &  $-10.45\pm 0.03$ &
                                $6.45\pm0.08$&
                                $0.45\pm0.01$
                                &$0.126$   &  $0.002\pm 0.004$    &       $7.88\pm0.06$      &   $5.68\pm 1.02$   &   $-10.35\pm 0.12$& $7.43\pm0.19$&$0.34\pm0.07$  \\
                                & S-Pos. &   $0.005$    &  $0.026\pm 0.003$    &    $8.53\pm 0.04$         &    $3.62\pm 0.84$  &   $-10.52\pm 0.14$& 
                                $6.84\pm0.46$&
                                $0.35\pm0.06$
                                &$0.029$    &  $0.030\pm 0.007$   &     $7.64\pm 0.20$        &   $5.26\pm 0.80 $   &  $-10.17\pm 0.19$ & $7.35\pm0.40$ &$0.35\pm0.15$ \\
                                \hline
\end{tabular}
}
\end{table*}

%%%%%%%OLD paper%%%%%%%%%%%%%%%%%%%%%%%%%%%%%%%%%%%%%%%%%%%%%%%%%%%%%%%%%%%%%%%%

\begin{figure*}
\resizebox{16cm}{!}{\includegraphics{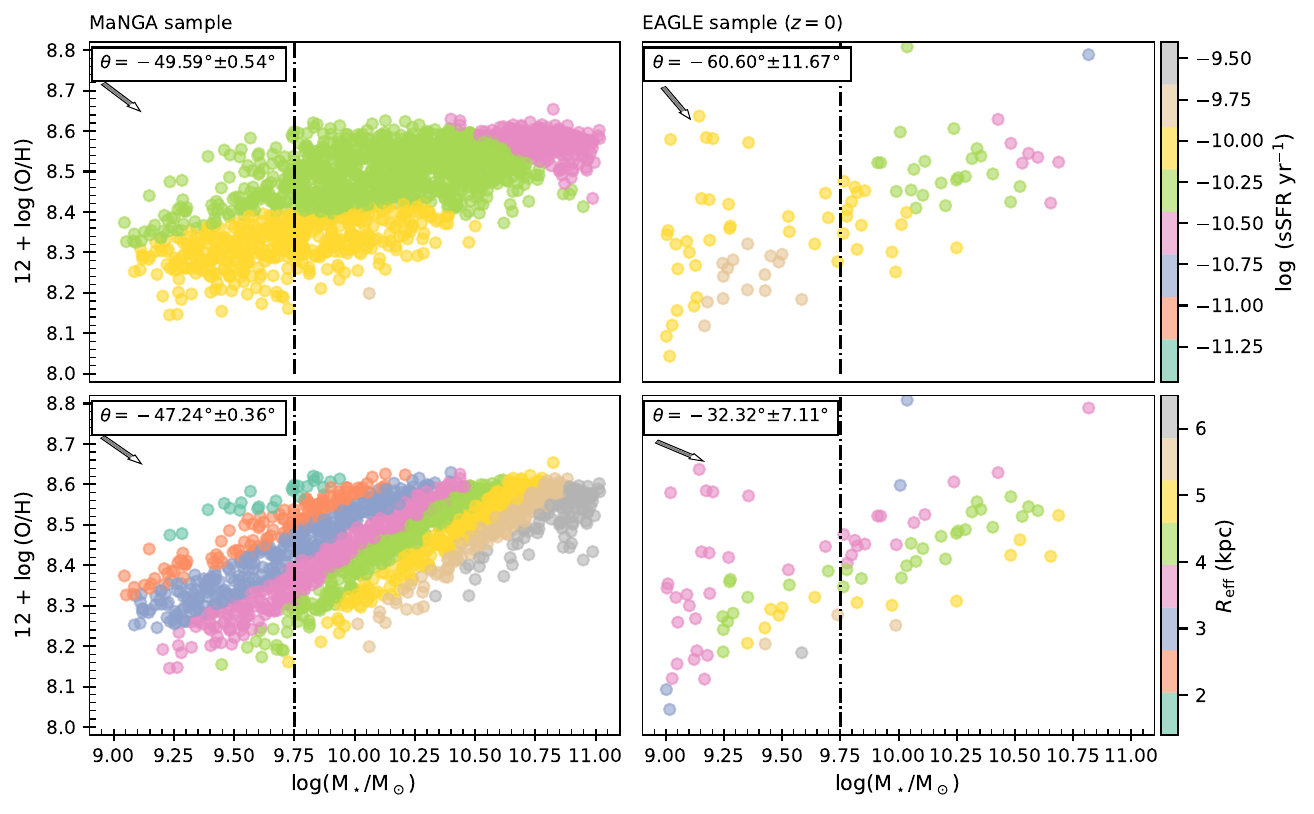}}
\caption{Mass-metallicity relation for the MaNGA (left panel) and \eagle~(right panel) samples coloured by sSFR (upper panels) and \Reff~(lower panels). To reveal the correlation between the properties, the Locally Weighted Scatterplot Smoothing (LOESS; \citealt{Cappellari2013}) technique was applied in all the panels. For low-mass galaxies (the dash-dotted black line denotes $\mstar  = 10^{9.75}~\Msun$) a correlation angle $\theta$ was computed using PCC analysis \citep[Partial Correlation Coefficients,][]{bluck2020}. This angle denotes the direction in which the third dependency of each panel grows. \eagle~galaxies follow the observational trends for both parameters, with a global trend for galaxies with high metallicity to have lower star-forming and have smaller \Reff~for $\mstar \lesssim 10^{10}~\Msun$,  at a given stellar mass. Higher masses tend to have lower sSFR and to be more extended. The errors on $\theta$ were estimated by applying a bootstrap technique.}
\label{maps_ssfr_reff}
\end{figure*}

\section{Analysis and Results} \label{sec:analysis}

%The scatter in the MZR has been linked to different parameters such as sSFR, SFR  and gas fraction as discussed in the Introduction. 
In this section, we first investigate the MZR for galaxies exhibiting different metallicity gradients. Figure~\ref{mzrgrad} presents the MZR for both MaNGA and \eagle~samples in bins of \grad, categorised by the metallicity gradient intervals. The solid black line represents the main MZR relation defined by the  MaNGA sample. In order to better compare the behaviour of both samples on the MZR, we re-normalised the gas-phase metallicity of the \eagle~galaxies, which are known to be systematically more metal-rich compared to the MaNGA galaxies \citep{bahe2017}. In Appendix~\ref{sec:samples} we explained this in more detail (see also Fig.~\ref{eagle_norm}). % We calculated the difference between the mean metallicity value of \eagle~ and MaNGA galaxies at $\mstar\sim 10^{10}\Msun$. Then, we subtracted that quantity to all the metallicities of \eagle galaxies, decreasing the offset between the global medians of both samples, as shown in Fig.~\ref{eagle_norm}. %The final distribution of our MaNGA and \eagle~ samples in the three mass ranges is showed in this figure.

As can be seen from Fig.~\ref{mzrgrad} galaxies with weak metallicity gradients follow a similar relation to the full sample as they constitute the dominant population. On the other hand, galaxies with strong positive gradients are consistently less enriched at a given stellar mass, while those with strong negative gradients are more enriched for $\rm M_{\star} \lesssim 10^{10} \Msun$. For higher masses, the three subsamples tend to converge to a comparable level of enrichment, albeit with those possessing positive gradients located systematically below the full MZR. Moreover, it is worth mentioning that galaxies with strong positive gradients are less prevalent among intermediate and high mass galaxies.
%\eagle~ galaxies show  similar to those found in the MaNGA sample. Galaxies with $\rm M_{\star} \lesssim 10^{10} \Msun$ exhibit the same trend, with those having steep positive gradients showing lower levels of enrichment and those with strong negative gradients, being more enriched. For high stellar masses, there seems to be a convergence to approximately similar levels of enrichment, 
We also notice that there is a large dispersion in the \eagle~samples, which is partly due to the lower number of members in these subsamples, particularly at the high-mass end, compared to the MaNGA subsamples.

The scatter in the observed MZR has been shown to have a secondary dependence on  sSFR and \Reff, in the sense 
 that  galaxies with higher sSFR tend to have lower levels of enrichment and to be more extended, at a given stellar mass \citep{ellison2008}. As expected the MaNGA sample is consistent with this trend, as shown in the left panels of Fig. \ref{maps_ssfr_reff}. 
To reveal the correlation between the properties, the Locally Weighted Scatterplot Smoothing (LOESS; \citealt{Cappellari2013}) technique was applied to the MZR as a function of sSFR and  \Reff. For low-mass galaxies, we also computed a correlation angle, $\theta$, performing a PCC analysis (see Appendix \ref{sec:PCC}). This angle reflects how dominant a variable is on a plane formed by two other variables.
Our \eagle~sample shows a comparable trend to that defined by MaNGA galaxies, albeit with larger dispersion \citep[see also][who analysed a larger sample of \eagle~galaxies and found similar results]{zenocratti2022}.  Although the distributions are noisier due to the smaller number of galaxies in the simulated sample, the trends are globally reproduced as quantified by $\theta$ in Fig.~\ref{maps_ssfr_reff}. As mentioned before we note that low-mass galaxies  with low sSFR are not present in the \eagle~sample probably because of our strict selection criteria (see also Appendix~\ref{sec:samples}).
%We note that there is a population of galaxies with small \Reff, high sSFR, and low metallicity in both observational and simulated samples, which we will discuss in the next sections.

\subsection{ Metallicity gradients, the specific star formation and the stellar surface density}
\label{sec:surface}

%\begin{figure*}
%\resizebox{16cm}{!}{\includegraphics{Figuras/Fig3.pdf}}
%\caption{
%Median sSFR as a function of \sigmastar~for galaxies  in the MaNGA (thin lines and symbols) and MaNGA-1 subsamples (thick lines) with strong negative (orange triangles and lines), weak (purple crosses and lines) and strong positive (blue squares and lines) \grad. The top panels show the frequencies of galaxies with different \grad~as a function of \sigmastar (fraction of each gradient type per bin of \sigmastar). Error were estimated by using a bootstrap technique. }
%\label{ssfr_sigmastar}
%\end{figure*}

In order to relate the secondary dependence of MZR on \grad~found in Fig.~\ref{mzrgrad} with the star formation activity and the size,
 we investigate the relationship between the sSFR and \sigmastar, defined as $\mstar/$\Reff$^{2}$, for galaxies with different \grad~in each mass interval. 
\sigmastar~provides a metric for the stellar compactness of a galaxy, indicating that, at a specific stellar mass, there is a spectrum of potential sizes. In Table~\ref{table1} the medians of the parameters analysed are displayed \footnote{In order to increase the number of \eagle~galaxies, the simulated sample is extended by including those with $z \leq 0.18$.
Although this redshift range is wider than that of MaNGA, it allows us to compensate for the low volume size and render a more robust analysis. We confirmed that the dependence on \grad~is still present although the full MZR is shallower \citep{derossi2017}.
}.

\begin{figure*}
%\resizebox{16cm}{!}{\includegraphics{Figuras/Fig3a.pdf}}
%\resizebox{16cm}{!}{\includegraphics{Figuras/Fig3b.pdf}}

\resizebox{16cm}{!}{\includegraphics{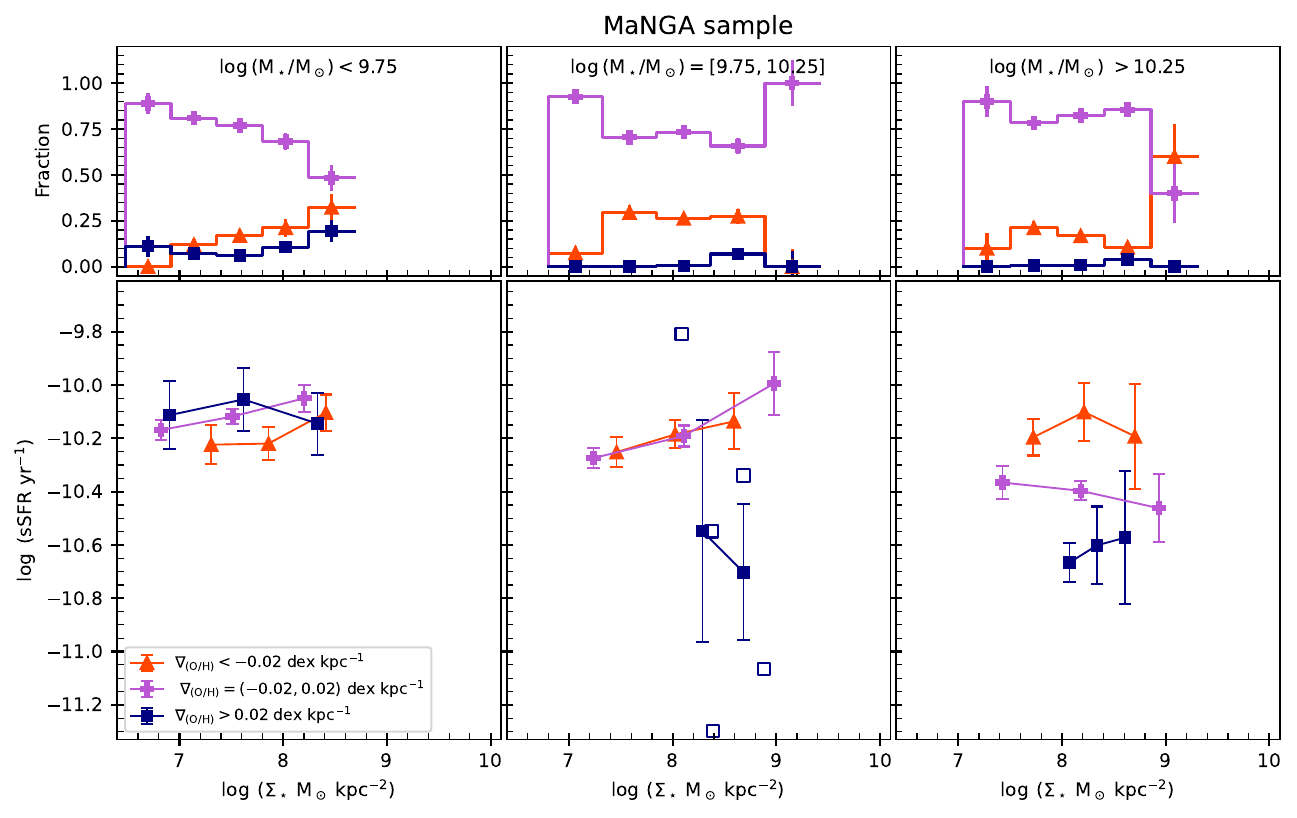}}
\resizebox{16cm}{!}{\includegraphics{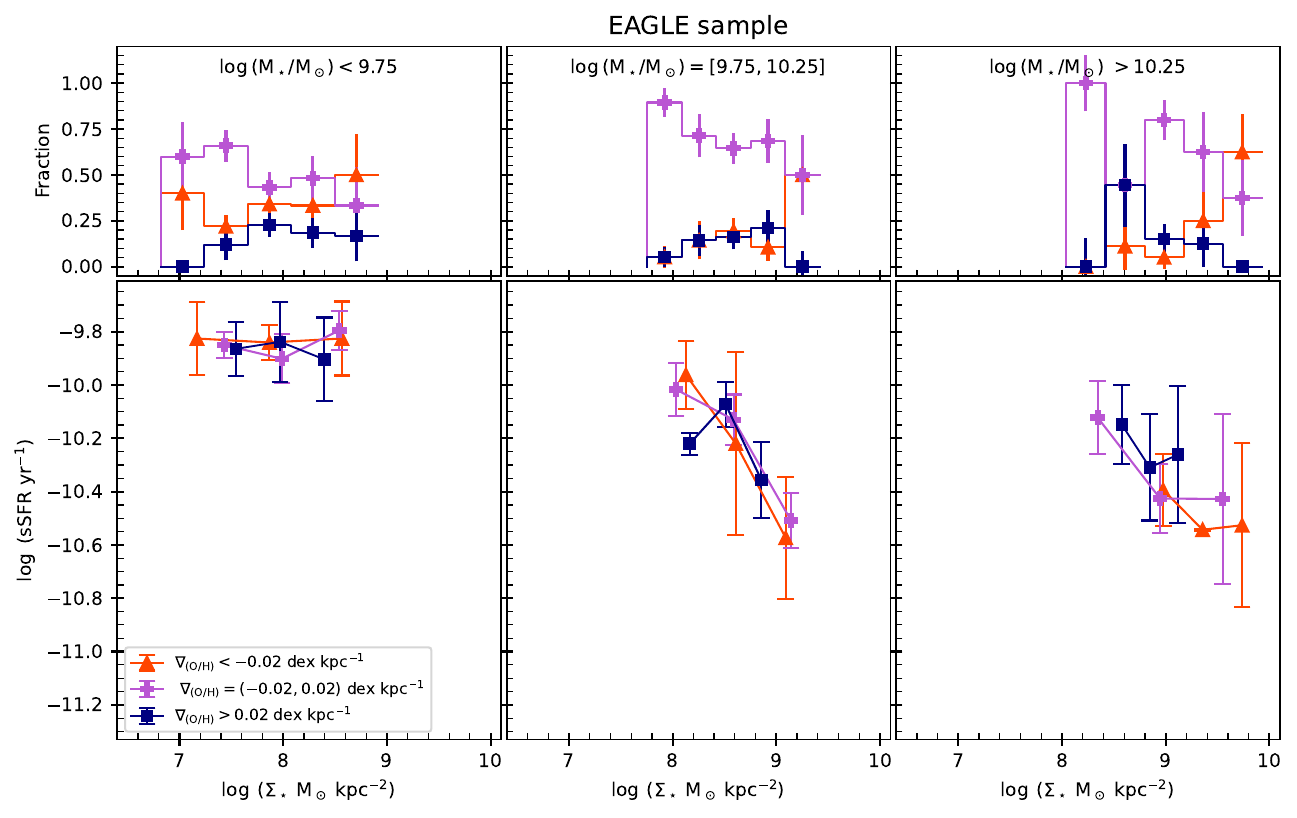}}
%\resizebox{8.cm}{!}{\includegraphics{Figuras/pdf/l25z0-mzr-grads_0.02.pdf}}
\caption{{\it Upper panel}: Median sSFR as a function of \sigmastar~for galaxies  in the MaNGA  with strong negative (orange triangles and lines), weak (purple crosses and lines) and strong positive (blue, filled, squares and lines) \grad. Individual values (open blue squares) have been added when number of galaxies in the sample is below 5).
{\it Lower panel}: Median sSFR as a function of \sigmastar~for galaxies  in the \eagle~sample  with strong negative (orange triangles and lines), weak (purple crosses and lines) and strong positive (blue squares and lines) \grad.  All error bars were estimated by using a bootstrap technique.The top small panels display the fraction of galaxies per bin \sigmastar 
~in each of the subsmaples  a function of \sigmastar.
}
\label{ssfr_sigmastar}
\end{figure*}

The upper panels of Fig.~\ref{ssfr_sigmastar} show the relationship between sSFR and \sigmastar~for galaxies in the three defined stellar mass intervals and for the three defined gradient subsamples for the MaNGA galaxies (solid lines and symbols). %To improve the statistics, we also include the MaNGA-1 sample (thick dashed lines), for which the abundances have been estimated using the same metallicity indicator, O3N2. 
As can be seen  in this figure, low-mass galaxies (left panel) exhibit a high level of sSFR regardless of \grad, with no  clear trend present between \sigmastar~and sSFR, within the corresponding bootstrap errors.

However, intermediate-mass galaxies exhibit different behaviours depending on the metallicity gradient. As depicted in the middle panel of Fig.~\ref{ssfr_sigmastar}, galaxies with strong positive gradients are systematically less star-forming and more compact. On the other hand, galaxies with weak and strong negative gradients continue to form stars actively. %Galaxies in MaNGA-1 determine trends consistent with those of the MaNGA sample, although the former reveals that more concentrated galaxies with strong negative and weak gradients might have slightly lower sSFR.

In the high stellar mass interval (right panel of Fig.~\ref{ssfr_sigmastar}), galaxies tend to be more quenched as their \grad~becomes shallower and more positive. However, galaxies with strong negative gradients continue to form stars actively, regardless of their \sigmastar. These galaxies show only a small decrease in their sSFR, of at most $\sim -0.10~\rm dex$, as their stellar mass increases from the low to the high end (see Table~\ref{table1}). On the other hand, galaxies with positive \grad~exhibit lower sSFR levels for all \sigmastar~values and the decrease of sSFR is $\sim -0.40~\rm dex$ from low to high mass galaxies (Table~\ref{table1}). 

The upper, left panel of Fig.~\ref{ssfr_sigmastar} reveals that  compact, low stellar mass galaxies tend to have higher frequencies of both strong positive and negative \grad. Conversely, galaxies with weak gradients are more frequent in less concentrated galaxies which suggest they tend to have a large frequency of extended and disc-dominated galaxies. The trend with size can be also seen in Table~\ref{table1}, which displays the median sizes for galaxies in each mass interval and \grad~subsample.

%We can appreciate that there is a strong correlation between metallicity gradients and \sigmastar~ \citep[see also][]{boardman2023}.

The upper, middle and right panels of Fig.~\ref{ssfr_sigmastar} show that for galaxies with masses higher than $10^{9.75}~\Msun$, weak gradients are also more common and evenly distributed at all \sigmastar. The presence of strong types becomes less significant for increasing galaxy mass. Galaxies exhibiting strong positive gradients are rare and tend to be more concentrated within a given mass interval, whereas galaxies with strong negative gradients are more evenly distributed as a function of \sigmastar~in high mass galaxies than in low-mass systems.
These trends, along with those found for sSFR as a function of \sigmastar, persisted for most of the metallicity calibrators that we tested, as shown in Fig.~\ref{sigmastarcalibrators}.

In summary, low-mass galaxies exhibit high sSFR irrespective of \grad~and do not show any trend with \sigmastar, in agreement with the findings for MaNGA galaxies
However, the dependence of sSFR on \sigmastar~by \eagle~galaxies is significantly different with respect to the MaNGA's trends for intermediate and high mass galaxies. Both intermediate and high-mass galaxies in \eagle~exhibit a systematic decline in sSFR as \sigmastar~increases, independently of \grad.

From Fig.~\ref{ssfr_sigmastar}, it can be seen that  the quenching start to manifest clearly for both observational and simulated galaxies  within the intermediate mass interval. This agrees with the mass range where AGN feedback is expected to act more efficiently in \eagle~simulations \citep[e.g.][]{vanloon2021}. The transition mass between the regimes dominated by SNe feedback and that dominated by AGN feedback is expected to be around $10^{10}~\Msun$.
The lack of dependence of the sSFR-\sigmastar~relation on \grad~in the \eagle~galaxies might suggest that the AGN feedback model used could be more efficient at decreasing the star formation activity by heating the gas than mixing or expelling the chemical elements in the star-forming regions in the simulated discs.  Hence, this tension could, indeed, contribute to improve the subgrid physics related with star formation and feedback mechanisms. Our findings suggest that the diminishing of sSFR as a function of the stellar compactness can be also linked to the characteristics of the distribution of chemical abundances in the discs, encoded by the metallicity gradients.

\subsection{The impact of strong accretion history in \eagle~galaxies}

In this section, we  examine the impact of strong accretion events on the location a galaxy occupies on the MZR in the context of the results reported in Fig.~\ref{mzrgrad}. As mentioned before, accretion events could be gas inflows, minor or major mergers. We adopt the parameter
$\tau_{25}$ to quantify the occurrence of the latest strong event that caused a significant change of stellar mass as explained in Section \ref{sec:eaglesample}.
%which measures the time elapsed since the last major increase of stellar mass (i.e. more than 25 per cent increase of stellar mass between two available snapshots), regardless of processes that could caused it (e.g. gas inflows or mergers).
%while $\tau_{\rm MM}$ measures the time since the last major merger.

\begin{figure}
%\resizebox{16cm}{!}{\includegraphics{Figuras/mzr-ssfr-reff-dt-loess+pcc.pdf}}
\resizebox{8cm}{!}{\includegraphics{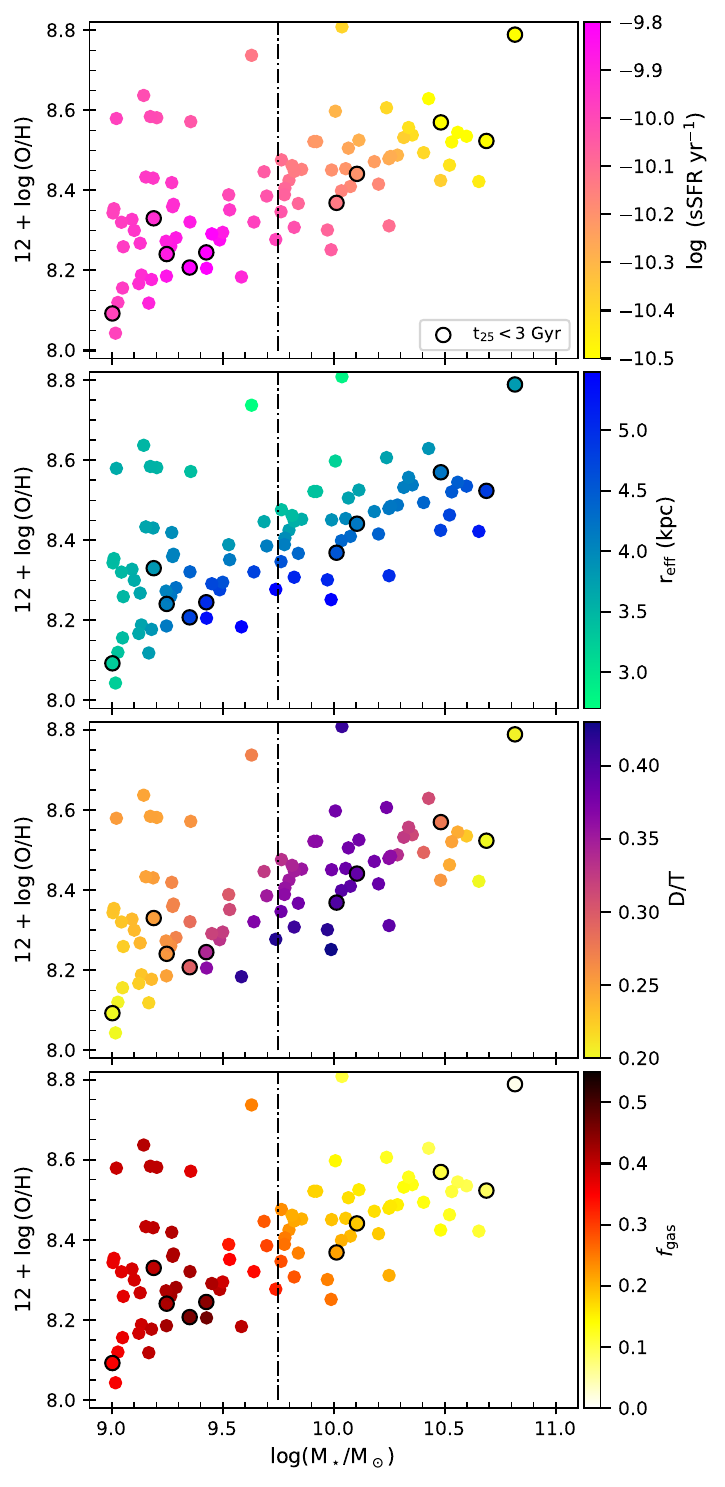}}
\caption{Mass-metallicity relation coloured by specific star formation rate, sSFR (upper panel), effective radius, \Reff~(second panel), disc-to-total stellar mass ratio, D/T (third panel) and gas fraction, $f_{\rm gas}$ (lower panel). To reveal the correlation between the properties, the Locally Weighted Scatterplot Smoothing (LOESS; \citealt{Cappellari2013}) technique was applied in all the panels. Galaxies with t$_{25}$ lower than $3$ Gyr have been  highlighted  with a black circle. 
The dotted-dashed line denotes $\mstar = 10^{9.75}~\Msun$.
%caption option 2:
%Mass-metallicity relation coloured by specific star formation rate, sSFR (upper panel), effective radius, \Reff~(second panel), disc-to-total stellar mass ratio, D/T (third panel), gas fraction, $f_{\rm gas}$ (fourth panel) and metallicity gradien, \grad~(lower panel). To reveal the correlation between the properties, the Locally Weighted Scatterplot Smoothing (LOESS; \citealt{Cappellari2013}) technique was applied in all the panels. Galaxies with t$_{25}$ lower than $6$ Gyr have been  highlighted  with a black circle. The dotted-dashed line denotes $\mstar = 10^{9.75}~\Msun$.
}
\label{fig: mzr-loess-pcc}
\end{figure}

In order to better understand the relationship between the MZR and the recent strong accretion history of galaxies, we have computed the MZR as a function of sSFR, \Reff,  D/T and $f_{\rm gas}$, as shown Fig.~\ref{fig: mzr-loess-pcc}. Those galaxies which had a main accretion event in the last 3~Gyr, i.e. $\tau_{25} < \rm 3~Gyr$,  are highlighted by black circles. This value of $\tau_{25}$ is taken from T22 where an excess of strong gradients was detected from galaxies with strong accretion events within this time.
%is the approximately $0.5\times(\tau_{25}^{\rm max} - \tau_{25}^{\rm min})$.
%Decreasing this threshold does not affect the results but they become noisier as the number of events decreases while increasing it makes the signal disappeared as expected.  
%, while the right panels show the full \eagle~sample.
We applied the LOESS algorithm \citep{Cappellari2013}  in all the panels. 
From this figure, we can see that galaxies with $\mstar \lesssim 10^{9.75}~\Msun$ (dashed vertical line) and recent accretion events are mostly high star-forming galaxies. Those with low-metallicity tend to be more extended and more disc-dominated. However, at the lowest mass end, there are galaxies which are mostly small and spheroidal dominated (see also Table\ref{table1}.

Massive galaxies with recent accretion have similar  sSFR and \Reff~ than the rest of the populations as shown by the first and second panels of Fig.~\ref{fig: mzr-loess-pcc}. However, there is a weak trend to have higher metallicity at a given stellar mass (black encircled symbols). In fact, some \eagle~ massive galaxies with recent accretion have strong negative \grad~as pointed out by T22. This could be explained by the fact that  massive systems might be more stable against external forces and, hence, the accreted low-metallicity material can settle onto a disc, contributing to steepen the negative metallicity gradients \citep{collacchioni2019}. They could also have more prominent bulges, which can also provide stability to the systems.
From the third panel of Fig.~\ref{fig: mzr-loess-pcc} we can see that more disc-dominated galaxies populate the intermediate mass region and that for low mass galaxies, there is a population of more disc-dominated systems with low metallicity, which are also more extended and have low $\tau_{25}$, as displayed in the previous panel. The fourth panel shows the trend with $f_{\rm gas}$ which show the expected trend for most the low mass galaxies with recent accretion events to have larger gas fractions. Massive galaxies are more gas-poor than low mass galaxies as expected. 

Hence, at a given stellar mass, high metallicity galaxies could be more passively evolving and show negative metallicity gradients or could have been part of a major merger event, which contributes with low metallicity gas that reinforces the negative slopes and/or triggers strong star formation activity. The latter is found to be more common in low stellar mass galaxies and the former, in massive ones. Galaxies with positive metallicity gradients are preferentially found at lower metallicity for a given stellar mass (Fig.~\ref{mzrgrad}), and this could indicate that they have experienced inward radial inflows or strong SNe outflows. These mechanisms dilute the metallicity in the inner region as studied in previous works \citep[e.g.][]{rupke2010, perez2011,dimatteo2013,moreno2019}.

Although we find some trends with morphologies,  we did not find clear differences in median of the morphological indicators as can be seen from  Table~\ref{table1}. This aspect will be delayed to future work.

\subsection{The impact of AGN feedback in galaxies}

\begin{figure}
\resizebox{8cm}{!}{\includegraphics{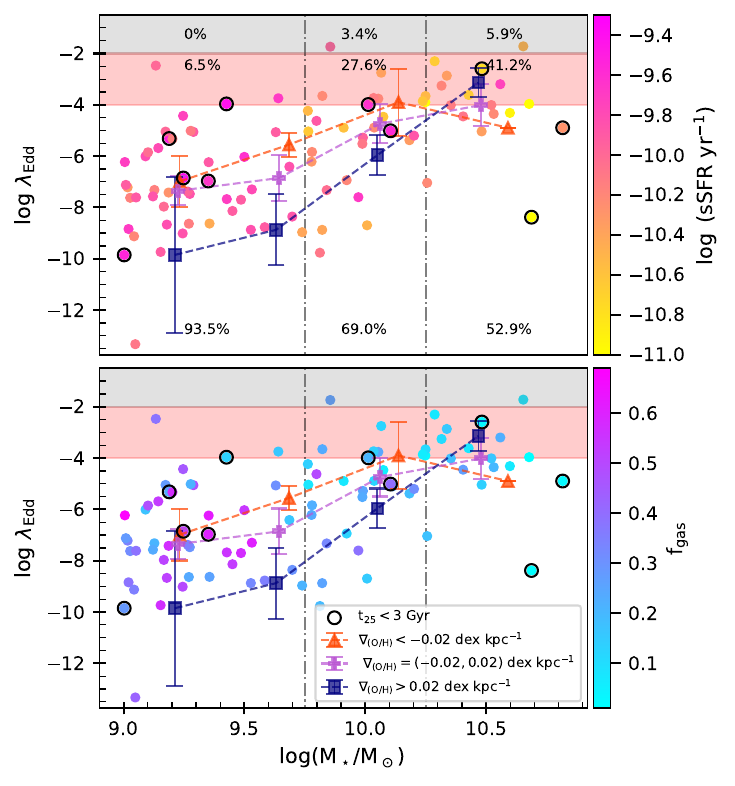}}
\caption{Eddington ratios, $\lambda_{\rm{{Edd}}}$, as a function of $\mstar$, coloured by the sSFR (upper panel) and the gas fraction, $f_{\rm{gas}}$ (lower panel). The median trends estimated for galaxies with strong negative (orange triangles and dashed lines), weak (purple crosses and dashed lines) and strong positive (blue squares and dashed lines) metallicity gradients are also shown. The grey shaded area represents the region where galaxies are considered as AGN, while the pink shaded area represents the region were galaxies follow the ADAFs (Advection Dominated Accretion Flows) regime. Both panels contain the threshold adopted to define out stellar mass subsamples (dash-dotted line). Galaxies with recent massive mergers or accretion are denoted by a black enclosing circle (i.e. t$_{25} < 3$ Gyr).   Additionally, the upper panel shows the percentage of galaxies within each accretion regime per mass bin.
%the SSD accretion regimen zone for each stellar mass subsample.
}
\label{fig: mzr-edd-ssfr-fgas}
\end{figure}

In this section we analyse the possible impact of AGN feedback in the relations analysed in Fig.~\ref{ssfr_sigmastar}.
For this purpose we analyse $\lambda_{\rm{{Edd}}}$  and $M_{\rm BH}$. In the case of the MaNGA sample we only have an estimation of $M_{\rm BH}$. 

%As mentioned above, we will assume  galaxies with $\rm log (\lambda_{\rm{{Edd}}})> -2$ as AGN systems and galaxies with $-2 < \rm log (\lambda_{\rm{{Edd}}})< -4$ as systems in the ADAF regime   \citep[][]{narayan1994, abramowicz1995}.
%We also assess possible dependence of the properties on  $M_{\rm BH}$ because the estimation of $\lambda_{\rm{{Edd}}}$ could be more uncertain as they involve instantaneous estimations of the parameters \citep{rosas-guevara2015}.

In  Fig.~\ref{fig: mzr-edd-ssfr-fgas}, the  $\lambda_{\rm{{Edd}}}$ as a function of $\mstar$ %as a function of the sSFR and $f_{\rm gas}$ 
is displayed for galaxies in the three different \grad~intervals. The upper panel also shows the dependence on sSFR while the lowest ones,  on $\rm f_{\rm gas}$. As can be seen there is a clear correlation in each of the \grad~ intervals, so that the level of activity increases for increasing $\mstar$. We recall that more massive galaxies are expected to have more massive SMBHs. However, from this figure we can see that there are very few  AGNs according to the adopted classification (see Section 2). Galaxies with intermediate and high masses also have lower star formation activity and lower gas fractions than low mass ones as expected. Only $6.5$ per cent of the galaxies in the low-mass interval has relevant accretion activity (ADAFs) and none hosts an AGN as expected.

We have also identified galaxies that have had a recent strong mass growth event quantified by $\tau_{25} < 3\rm~Gyr$ as in  Fig.~\ref{fig: mzr-loess-pcc}.  As noted in Fig. \ref{fig: mzr-edd-ssfr-fgas}, these galaxies  are preferentially low mass (5 over 10 galaxies). This supports the claim that low-mass galaxies are mainly regulated by SN feedback, probably triggered by strong accretions which induced starbursts as these systems are also gas-rich and have the higher sSFR.

 Interestingly, galaxies with different metallicity gradients occupy different regions in Fig.~\ref{fig: mzr-edd-ssfr-fgas}. Galaxies with strong negative gradients tend to be located above the median relation for galaxies with weak metallicity gradients, whereas galaxies with strong positive \grad~prefer to be clearly below it. Most of low mass galaxies are inactive as expected.  
 At a given stellar mass, there is a systematic increase of the $\lambda_{\rm{{Edd}}}$ from galaxies with strong positive to strong negative \grad. Because the star formation activity also increases in the same fashion, we speculate that the process feeding the star formation, i.e. galaxy mergers, could be also doing the same with the central BH. And, at a given $\lambda_{\rm{{Edd}}}$, galaxies with positive gradients tend to be more massive than the rest. The correlations remains when $\mstar$ is replaced by \sigmastar, so that more massive and compact galaxies show higher $\lambda_{\rm{{Edd}}}$ and at a given, \sigmastar, galaxies with strong positive gradients  have lower $\lambda_{\rm{{Edd}}}$. 

\begin{figure*}
\resizebox{16cm}{!}{\includegraphics{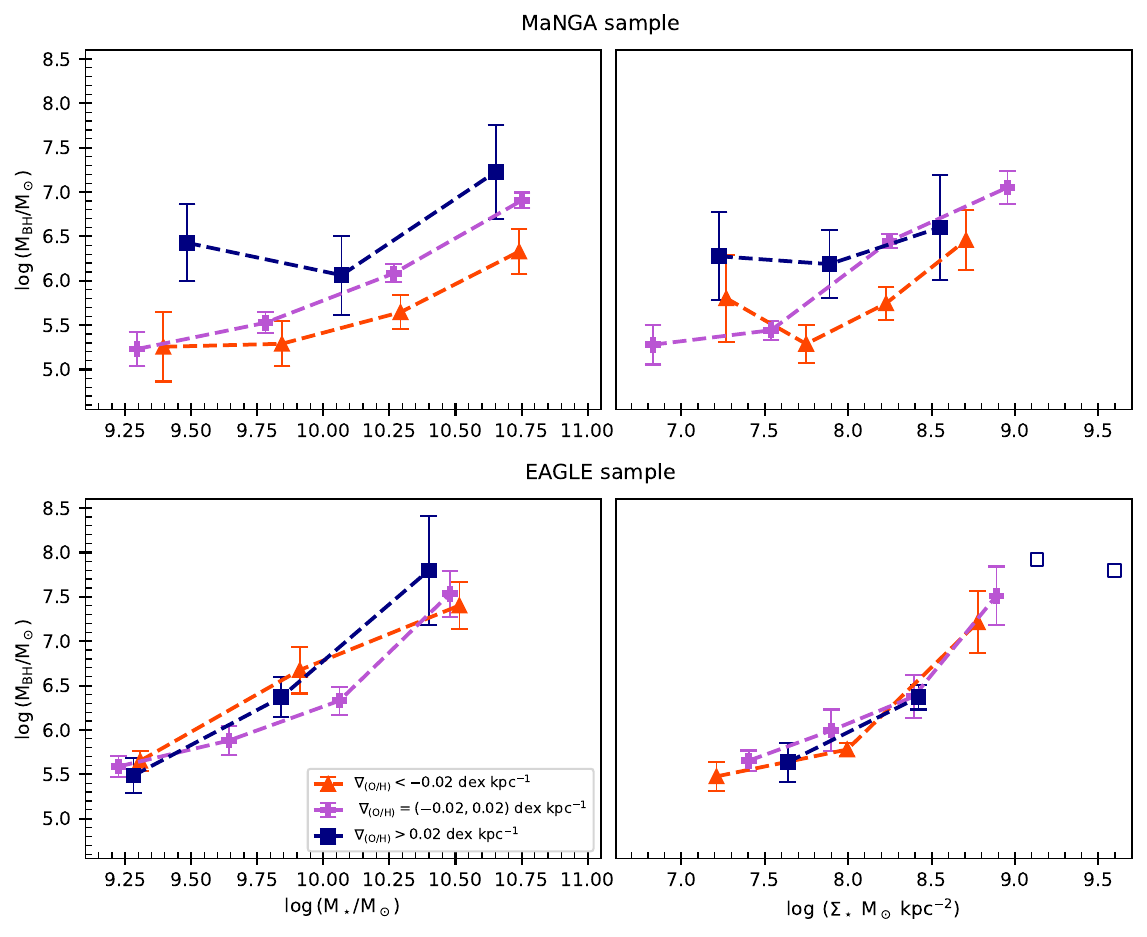}}

\caption{$\rm M_{\rm BH}$  as a function of $\mstar$ and \sigmastar~for galaxies in the MaNGA (upper panels) and \eagle~sample (lower panels) separated according to their metallicity gradients as displayed in the inset labels.}
\label{fig: MBH}
\end{figure*}

The $\rm M_{\rm BH}$ is known to be more stable and is the result of the BH growth along the history of formation of galaxies. Hence, we analyzed it as a function of both $\mstar$ and \sigmastar~for intermediate and high mass galaxies since they are potential affected by AGN feedback. Fig.~\ref{fig: MBH} displays the trends for the MaNGA (upper panels) and the \eagle~(lower panels) samples. We note that the \eagle~galaxies have more massive $\rm M_{\rm BH}$ at a given $\mstar$ or \sigmastar. However,
in this analysis we are only interested in assessing the relative trends between the relations for galaxies with different \grad~within each sample.

 At a given stellar mass, galaxies in MaNGA  have larger $\rm M_{\rm BH}$ if they have strong positive gradients. Indeed, there is a systematic decrease of the black hole mass from galaxies with strong positive, weak to strong negative gradients. As a function of \sigmastar~galaxies with strong negative gradients also tend to have smaller $\rm M_{\rm BH}$ at a given stellar mass and \sigmastar. Those with strong positive gradients tend to have more  massive BHs, which is consistent with having already experienced AGN activity which fed the BHs and diminish the star formation activity via AGN feedback as shown in Fig.~\ref{ssfr_sigmastar}. However, as a function of \sigmastar, there is no significant differences between the trends of galaxies with weak and positive \grad.

For the \eagle~sample, the correlations are also present but there is no clear difference between galaxies with different gradients as can be seen from the lower panels of 
Fig.~\ref{ssfr_sigmastar}. There are only indication that galaxies with strong gradients would tend to have larger $\rm M_{\rm BH}$ than those with weak ones at a given stellar mass. This could be due to the impact of mergers and strong gas inflows (see T22) which can feed the bulge in the \eagle~simulations. 

%From the MaNGA survey, we obtained that  galaxies with strong negative gradients tend to have smaller central black holes than galaxies with weak gradients,  at a given stellar mass; they are also actively forming stars.    These trends are not found in the \eagle~sample.

The results reported in this section could explain the behaviour of the \eagle~galaxies in the sSFR - \sigmastar~relation and their apparent independence on metallicity gradients, suggesting that the AGN feedback  might be  efficient at diminishing the star formation for more compact galaxies, but without strongly affecting the metallicity distribution in the star forming regions across the disc. Additionally, this could indicate the need to revise the modelling of both SN and AGN feedback.

\section{Conclusions}
\label{sec:conclusions}
We analysed galaxies from the MaNGA survey and from the \eagle~Project to understand the connection between the metallicity gradient, the global metallicity, and the stellar mass as a function of galaxy size, star formation and AGN activity. The ability to access the evolutionary history of the \eagle~galaxies enabled us to investigate the information stored in the metallicity gradients of galaxies in relation to their recent history of significant accretion.
The metallicity gradients were estimated for the star-forming discs within [0.5, 1.5]\Reff~for both observations and simulations. For the observations we performed several tests by using different metallicity calibrators, finding similar global results for most of them.

Our main results are the following:
\begin{itemize}
\item Our analysis shows that galaxies with different metallicity gradients tend to occupy different regions of the MZR at a given stellar mass. Galaxies with stellar masses lower than $\sim 10^{9.75}~\Msun$ and  strong negative gradients tend to be more chemically enriched than those with strong positive gradients. However, for higher mass galaxies, the level of enrichment converges to similar values regardless of \grad.

\item A clear relation between the sSFR and  the stellar compactness is identified, which varies with the metallicity gradients for MaNGA galaxies. 
In observations and simulations, low mass galaxies are found to be actively forming stars irrespective of their compactness and metallicity gradients. However, galaxies with strong positive and negative gradients tend to be more compact, i.e., higher \sigmastar, while those with weak metallicity gradients tend to be less concentrated.

For intermediate and high mass galaxies, we found differences between the trends in MaNGA and \eagle. In MaNGA, we observed a decrease in sSFR with increasing \sigmastar~for galaxies with positive gradients. % at a stellar mass of $\sim 10^{10} \rm M_{\odot}$.
 For galaxies with weak gradients, the sSFR decreases with increasing stellar mass, while for galaxies with strong negative gradients, the sSFR remains high regardless of \sigmastar. The strong negative gradient galaxies tend to be more disc-dominated, while the more quenched galaxies are both more concentrated and more dispersion-dominated.
%\eagle  galaxies reproduce the trend on the MZR and the fact that positive metallicity gradients are more frequent in low stellar mass galaxies and show a diversity of \sigmastar. 
However, simulated galaxies with $\mstar \gtrsim 10^{9.75}~\Msun$ are systematically less star-forming for increasing stellar concentration regardless of the metallicity gradient.This is at odd with our observational findings. 
%Also, globally \eagle galaxies in our sample tend to be more concentrated and the frequency of strong negative gradients in this mass range is not clearly enhanced as in observations, in part because \eagle~produce weaker gradients \citet{tissera2019} and the low number of members in our sample.

\item 
Galaxies in \eagle~with recent strong accretion tend to be more star-forming and have strong negative or positive gradients, and hence. The results of T22 show that  \eagle~galaxies recover their weak  metallicity gradients after  $1.4-2$ Gyr, on average. Hence, galaxies participating in this event could be temporally located at low or high metallicity before they recover a weak gradients and set on the main MZR. At this stage, the contribute to the scatter of the MZR and  could show signals of outflows at this stage .

\item  On the one hand, we find that \eagle~galaxies with strong positive gradients tend to show weaker BH accretion activity, i.e. lower  $\lambda_{\rm{{Edd}}}$, lower sSFR  and are more compact 
  than those with weak and negative metallicity gradients. This suggests that those simulated galaxies with strong positive \grad~might already passed their more active star-forming and probably AGN phases. 
  Galaxies with strong negative gradients are associated with strong $\lambda_{\rm{{Edd}}}$, larger $\rm M_{\rm BH}$, high sSFR and larger gas fractions. 
Regarding $\rm M_{\rm BH}$,  galaxies with strong negative and positive gradients show a weak trend to have larger $\rm M_{\rm BH}$ than those with weaker gradients at a given stellar mass.  But, the \sigmastar-$\rm M_{\rm BH}$ does not depend on the metallicity gradients.
  On the other hand, in the  MaNGA sample, galaxies with strong negative gradients tend to have lower $\rm M_{\rm BH}$ for a given $\mstar$ and \sigmastar~for $\mstar > 10^{9.75}$. Additionally, galaxies with strong positive gradients have systematically more massive $\rm M_{\rm BH}$ than the rest of the galaxies.

\end{itemize} 

%Overall, our findings indicate that comprehending the interplay among the level of enrichment, stellar surface density, and star formation activity across galaxy discs is crucial for understanding the scatter in the MZR.
Both observations and simulations agrees that, at low masses, the sSFR is high and shows no dependence on \sigmastar~or \grad, indicating a different star formation regulation compared to intermediate and high masses. The latter are expected to be more affected by AGN feedback compared to smaller galaxies, which are affected mainly by SN feedback. At these higher masses, in simulations the sSFR decreases systematically with increasing \sigmastar, suggesting a more efficient quenching mechanism in the most compact systems that we could link to the action of AGN feedback. The lack of dependence of these trends on \grad~in the simulations suggests that  AGN feedback model could be more efficient at quenching the star formation activity in the central regions than ejecting or mixing chemical elements in the star-forming regions across the discs \citep[see ][for a discussion of different feedback modelling]{chenarico2023}.

%Addhese findings show that both observed and simulated galaxies start to be quenched at about similar stellar mass. Hence, considering that AGN feedback might be  impacting on the regulating of the star formation in the simulated massive galaxies for increasing compactness,  our results suggest that AGN feedback could be  one of the main processes acting on quenching the star formation in observations as claimed in previous observational and numerical papers \citep[e.g.][and references there in]{belli2023}.  
%On the other hand, the lack of dependence of the trends between sSFR and \sigmastar~on \grad~in the \eagle~simulations provides a mean to improve the  subgrid models.  

Our results suggest differences between observations and simulations in the metallicity recycling and mixing history of massive galaxies with different metallicity gradients, in relation to their global parameters. This discrepancies could contribute to further understanding the physical processes that take place in galaxy formation.

\section*{Acknowledgements}
{We thank the anonymous referee for useful comments which contribute to improve this paper. We thank Jorge Barrera-Ballesteros for useful comments.
PBT acknowledges partial funding by Fondecyt-ANID 1200703/2020, ANID Basal Project FB210003 and ERIS Millenium Nucleus.  
YRG acknowledges the support of the
“Juan de la Cierva Incorporation” fellowship (ĲC2019-041131-I).
MEDR acknowledges support from {\it Agencia Nacional de Promoci\'on de la Investigaci\'on, el Desarrollo Tecnol\'ogico y la Innovaci\'on} (Agencia I+D+i, PICT-2021-GRF-TI-00290, Argentina).
This work was supported by the Science and Technology Facilities Council (STFC) astronomy consolidated grants ST/P000541/1
and ST/T000244/1. 
We acknowledge the use of  the Ladgerda Cluster (Fondecyt 1200703/2020).
This project has been  partially supported by the European Union Horizon 2020 Research and Innovation Programme under the Marie Sklodowska-Curie grant agreement No 734374.
This work used the DiRAC@Durham facility managed by the Institute for Computational Cosmology on behalf of the \href{www.dirac.ac.uk}{STFC DiRAC HPC Facility}. The equipment was funded by BEIS capital funding via STFC capital grants ST/K00042X/1, ST/P002293/1, ST/R002371/1 and ST/S002502/1, Durham University and STFC operations grant ST/R000832/1. DiRAC is part of the UK's National e-Infrastructure.

}

\section*{Data availability}
{The \eagle~simulations are publicly available; see \citet{mcAlpine2016}. The reported relations in this paper will be available upon request.}
\bibliographystyle{mnras}
\bibliography{bibliography}

\newpage
\appendix

\section{Additional Material}
\label{otherproperties}

\subsection{\eagle~and MaNGA samples}
\label{sec:samples}
\begin{figure}
\resizebox{8cm}{!}{\includegraphics{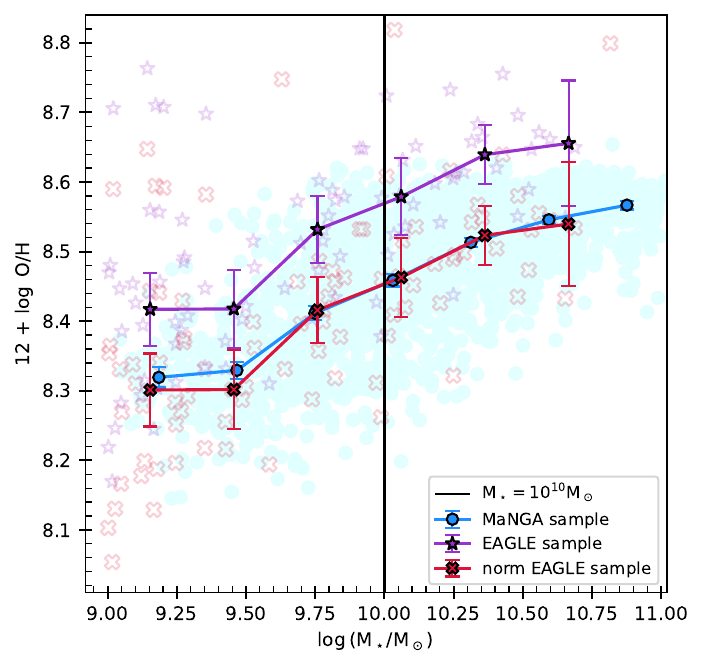}}

\caption{MZR for MaNGA (light-blue circles) and our original (purple stars) and normalized (red crosses) \eagle~samples. Medians are plotted with the raw data on the background. The black vertical line shows the limit $\mstar =10^{10}~\Msun$, which is the matching stellar mass between MaNGA and normalised \eagle~medians.}
\label{eagle_norm}
\end{figure}

In this section, we report some general considerations regarding our main samples, \eagle~and MaNGA. As described in Section \ref{sec:analysis}, we re-normalised the metallicity of \eagle~galaxies to match the enrichment level of galaxies in the MaNGA sample. We have taken as the normalisation point the median metallicity of MaNGA galaxies at $\mstar = 10^{10}~\Msun$, $\bar{\rm{x}}_{\rm{(O/H),\:manga}} \simeq 8.46$. Compared to the measured value in \eagle~galaxies, $\bar{\rm{x}}_{\rm{(O/H),\:eagle}} \simeq 8.58$, the difference   $\Delta\bar{\rm{x}}_{\rm{(O/H)}} \simeq 0.126$ was subtracted to all galaxies in the \eagle~sample. In the Figure \ref{eagle_norm}, we plotted the medians trends of the \eagle~sample before (purple stars) and after (red crosses) the re-normalisation, along with the median trend of MaNGA galaxies (light-blue circles). As can be seen in this figure, the re-normalised \eagle~sample reproduces the observable MZR with good accuracy. 

\begin{figure*}
\resizebox{15cm}{!}{\includegraphics{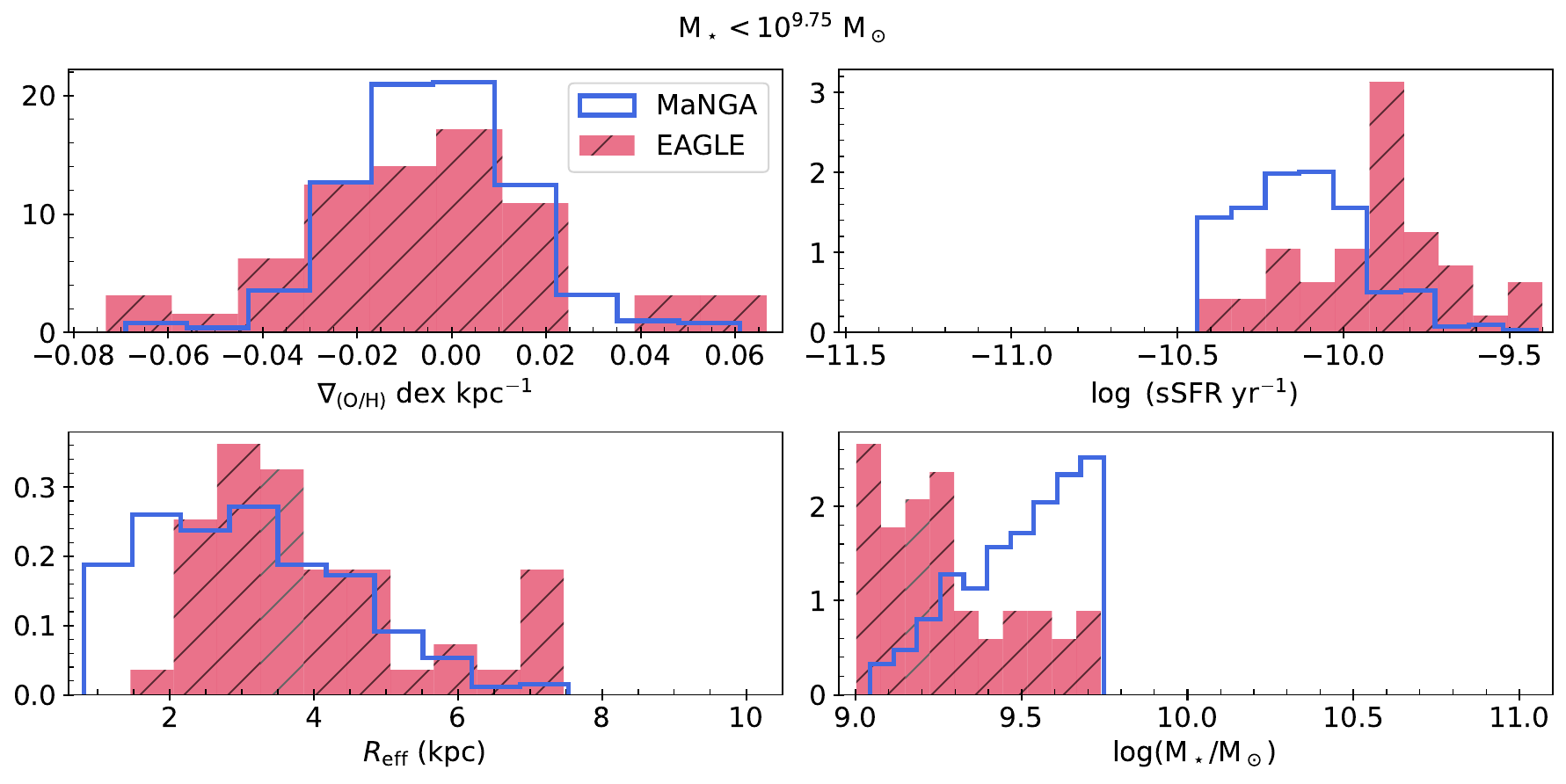}}
\resizebox{15cm}
{!}{\includegraphics{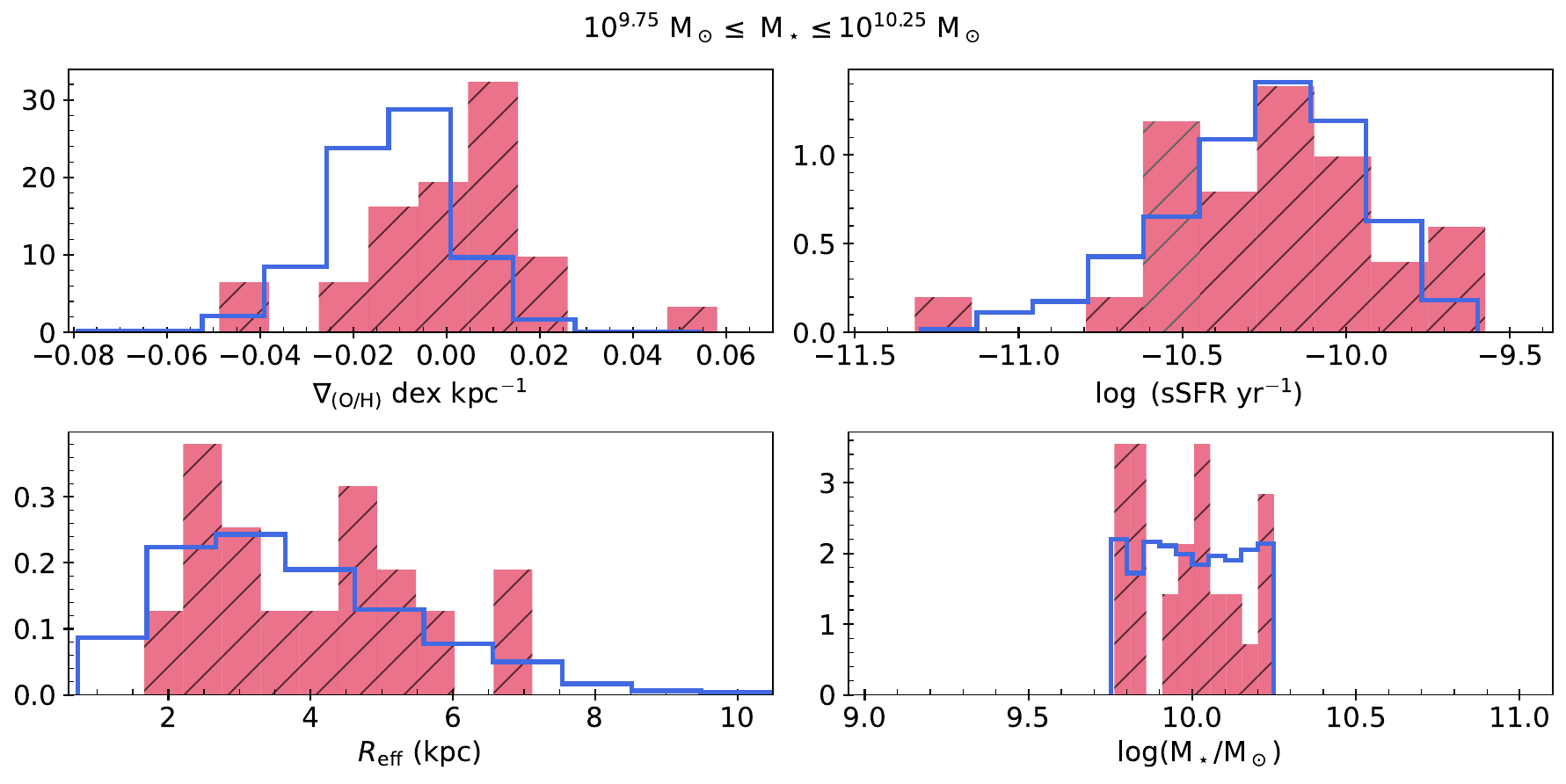}}
\resizebox{15cm}
{!}{\includegraphics{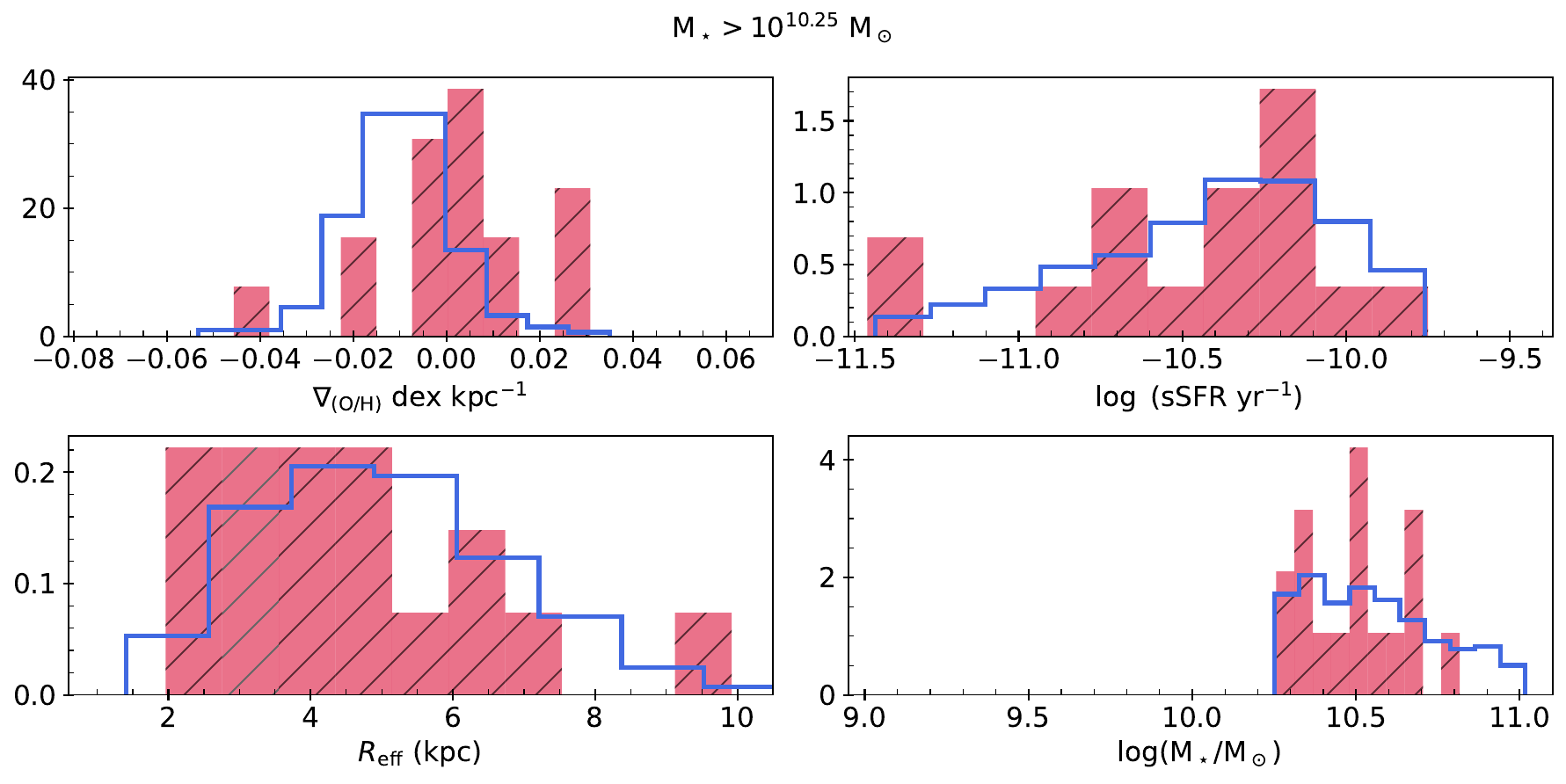}}
\caption{Comparative distribution of four main galactic properties, metallicity gradient \grad, specific star formation rate, sSFR, effective radius, \Reff, and stellar mass, $\mstar$, between our MaNGA (blue, solid lines) and \eagle~(pink, shaded areas) samples, displayed for each of the three defined stellar mass intervals as indicated by the labels in the figure.}
\label{fig: samples-histograms}
\end{figure*}

Having defined the re-normalised sample as our final \eagle~sample, in Fig.~\ref{fig: samples-histograms} we show the distribution of four key variables for the analysis carried out in this paper; gas-phase metallicity gradient \grad, specific star formation rate sSFR, effective radius \Reff, and stellar mass $\mstar$. Such distributions are deployed in mass ranges, for our \eagle~(pink striped regions) and MaNGA (blue lines) samples. In general, we can note that the distribution of these properties agrees globally between both samples, with some exceptions. At low masses, galaxies in the \eagle~sample are more star-forming, tend to be slightly more extended, and have a saturation at the lower end of the mass range, while galaxies in the MaNGA sample have masses closer to the higher end of the range. At intermediate and high masses, a higher level of agreement can be seen in the distribution of both samples, with a slight trend in \eagle~galaxies having more positive metallicity gradients. 

%In this Section, we show in  Fig \ref{maps_gradients}  the MZR in our MaNGA an EAGLE samples as a function of metallicity gradients in dex kpc$^{-1}$ and in dex \Reff$^{-1}$. 
%\begin{figure*}
%\resizebox{8cm}{!}{\includegraphics{Figuras/Heatmaps/heatmap-manga-mzr-grads.jpg}}
%\resizebox{8cm}{!}{\includegraphics{Figuras/Heatmaps/heatmap-eagle-mzr-grads.jpg}}
%\caption{MZR for our  MaNGA (left panel) and \eagle~(right panel) samples as a function of metallicity gradient \grad in units of dex %kpc$^{-1}$ (upper panels) and \Reff$^{-1}$ (lower panels). Normalizing by \Reff makes \grad consistently more negative in the overall %distribution, which is most evident in our MaNGA sample.}
%\label{maps_gradients}
%\end{figure*}

%\begin{figure}
%\resizebox{8cm}{!}{\includegraphics{Figuras/new figs/manga2-mass-size-grad.jpg}}
%%\resizebox{8.cm}{!}{\includegraphics{Figuras/pdf/l25z0-mzr-grads_0.02.pdf}}
%\caption{Mass-size plane for galaxies in our main and external MaNGA samples, colored by $\nabla_{\rm{O/H}}$ value normalized by $R_{\rm{eff}}$. From upper to lower panel, we plotted galaxies with strong negative \grad, weak \grad and strong positive \grad. All error bars were estimated by using a bootstrap technique.}
%\label{masssizecombined}
%\end{figure}

\subsection{Partial Correlation Coefficients}
\label{sec:PCC}

In Fig.~\ref{maps_ssfr_reff}, we used a correlation angle, $\theta$, to show the direction in which one variable (sSFR or \Reff) grows with respect to two others that form a plane (stellar mass and metallicity). This angle was computed from two measurements of a Partial Correlation Coefficient. A PCC reflects the correlation between two variables, $A$ and $B$, removing the effects of a third control variable $C$, and is defined as \citep{bait2017, bluck2019, bluck2020}

\begin{equation*}
    \rho_{A\cdot B|C} = \frac{\rho_{A\cdot B}-\rho_{A\cdot C}\cdot\rho_{B\cdot C}}{\sqrt{1-\rho_{A\cdot C}^2}-\sqrt{1-\rho_{B\cdot C}^2}},
\end{equation*}

where $\rho_{X\cdot Y}$ is the Spearman’s rank correlation coefficient between the variables X and Y. Then, the correlation angle $\theta$ can be defined as \citep{bluck2020b}

\begin{equation*}
    \theta = \tan^{-1}\left(\frac{\rho_{B\cdot A|C}}{\rho_{C\cdot A|B}}\right),
\end{equation*}
where $A$ represents the third variable to correlate with a plane constructed with $C$ ($x$-axis) and $B$ ($y$-axis).

Using this method, we computed PCC between  metallicity and sSFR, with stellar mass as a control variable, $\rho_{\rm sSFR\cdot(\rm{O/H})|\mstar}$, and the PCC between stellar mass and sSFR, with metallicity as a control variable, $\rho_{\rm sSFR\cdot\mstar|(\rm{O/H})}$ in both \eagle~and MaNGA samples to calculate $\theta$. Then we followed the same procedure using \Reff~instead of sSFR.

\section{Metallicity Calibrators}
\label{metallicitycalibrators}

\begin{figure*}
\resizebox{16cm}{!}{\includegraphics{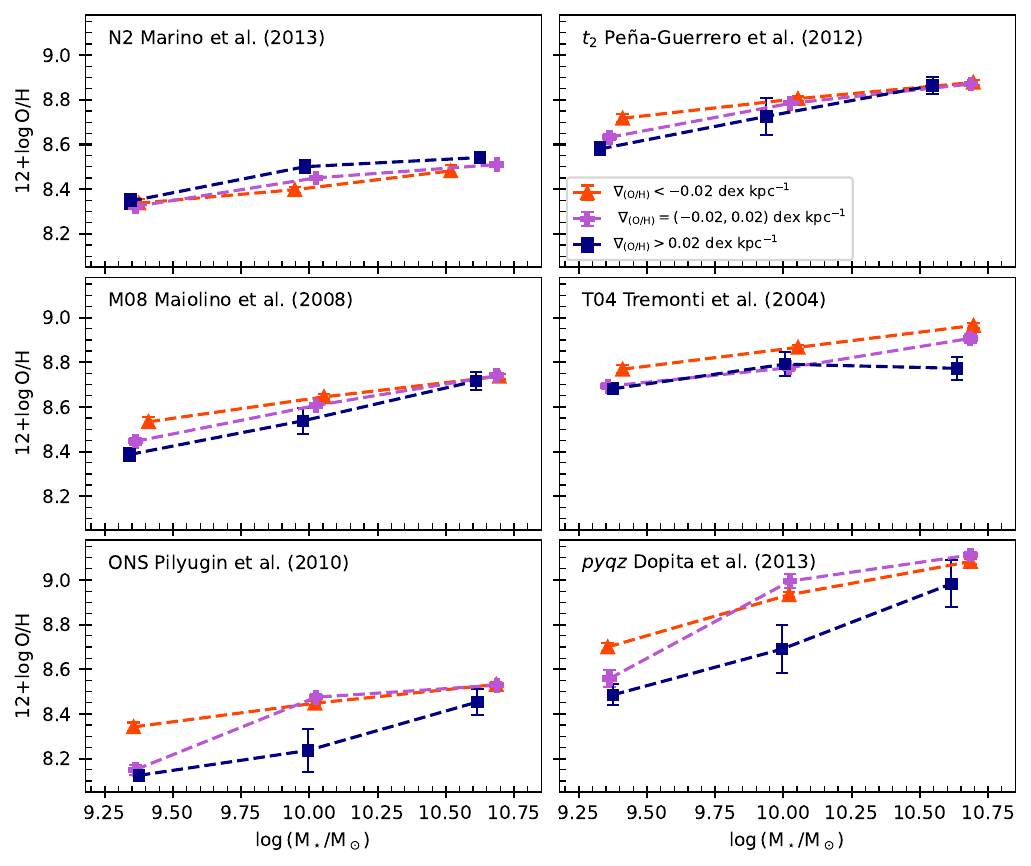}}
%\resizebox{8cm}{!}{\includegraphics{Figuras/pdf/l25z+-vsig-sigma.pdf}}
%\resizebox{8.cm}{!}{\includegraphics{Figuras/pdf/l25z0-mzr-grads_0.02.pdf}}
\caption{Dependence of MZR on metallicity gradients in our MaNGA sample. Comparison of the trends obtained from the six additional metallicity calibrators as mentioned in Section \ref{sec:data}. Orange, purple and blue shaded lines represent the median trends of galaxies with strong negative, weak and strong positive metallicity gradients, respectively. All error bars were estimated by using a  bootstrap technique.}
\label{mzrdifcalibrators}
\end{figure*}

In this Section, we estimated the MZR for galaxies in the three defined metallicity gradients by using different metallicity calibrators as shown in Fig.~\ref{mzrdifcalibrators}. Additionally, we performed the relations between sSFR and \sigmastar~as well as the distribution of galaxies with different \grad~as a function of \sigmastar~by using different calibrators.

%-------------------

Two of these calibrators follow a direct method to measure oxygen abundance, N2 \citep{marino2013} based on [NII] $\lambda6583$ and ONS \citep{pilyugin2010} which uses eight strong lines of oxygen, nitrogen and sulphur. Other two are based on pure photoionization models, the one used by \citet{tremonti2004} to explore the MZR, which is a modification of the classical R23 \citep{kobulnicky2004,rosalesortega2011} and depends only on oxygen line ratios, and the \textit{pyqz} proposed by \citet{dopita2013}, which uses combination of line ratios of  O2, N2, S2, O3O2, O3N2, N2S2 and O3S2. Among the two remaining, one is a mixed calibrator, M08 \citep{maiolino2008} also based on R23. The other is known as the $t_2$ correction proposed by \citet{peñaguerrero2012}, which considers an average of the abundances given by O3N2, N2, R23, ONS and a modified O3N2 described by \citet{perezmontero2009}.

%----------------------------

\begin{figure*}
\resizebox{16cm}{!}{\includegraphics{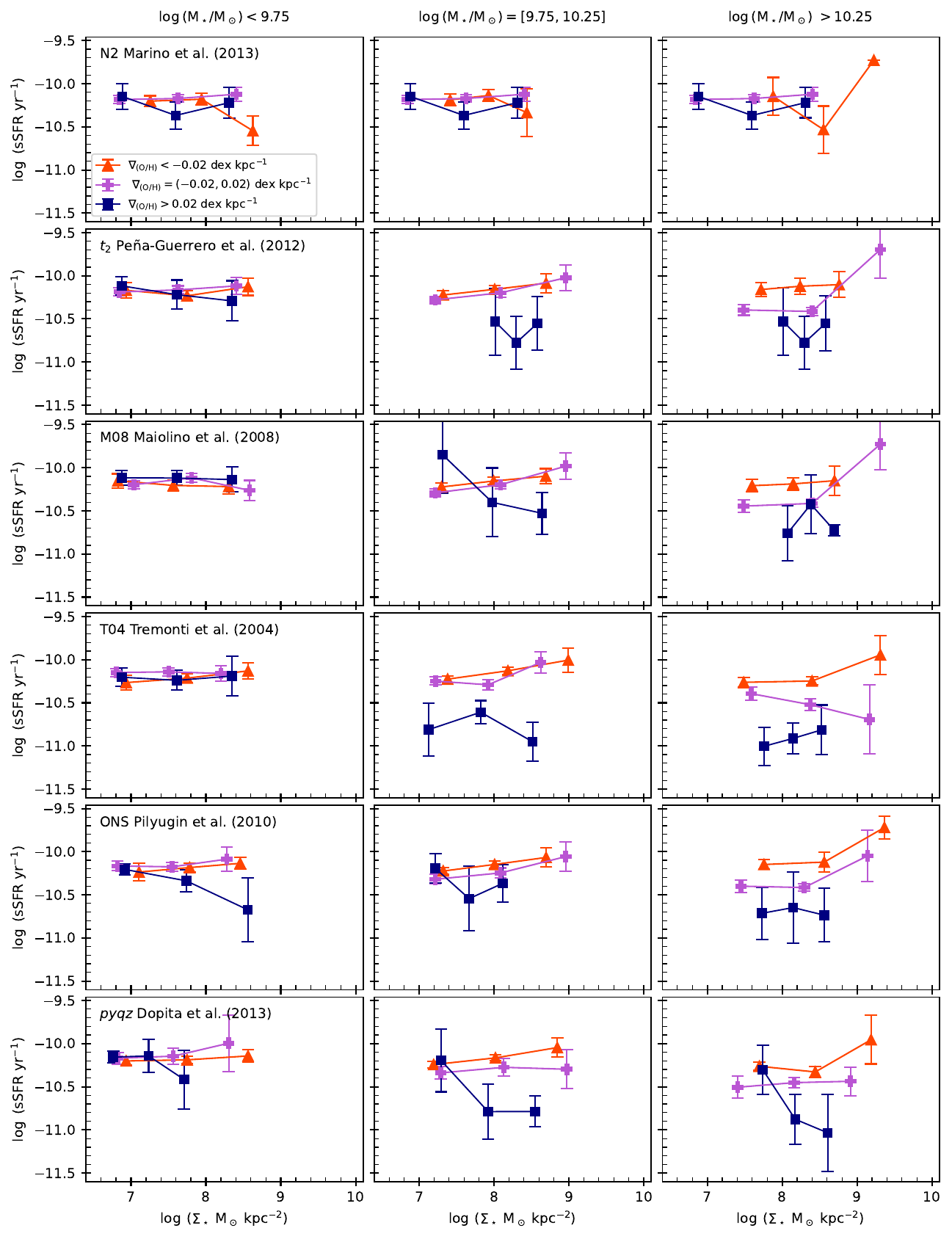}}
%\resizebox{8cm}{!}{\includegraphics{Figuras/pdf/l25z+-vsig-sigma.pdf}}
%\resizebox{8.cm}{!}{\includegraphics{Figuras/pdf/l25z0-mzr-grads_0.02.pdf}}
\caption{Specific star formation rate, sSFR, as a function of stellar mass surface density, \sigmastar, for our sample of MaNGA galaxies. Comparison of the trends produced from the six additional metallicity calibrators. Orange, purple and blue solid lines represent the median trends of galaxies with strong negative, weak and strong positive metallicity gradients, respectively. The sample was subdivided in low- (left panels), intermediate- (centrals panels) and high- (right panels) mass galaxies. All error bars were estimated by using a  bootstrap technique.}
\label{sigmastarcalibrators}
\end{figure*}

\begin{figure*}
\resizebox{16cm}{!}{\includegraphics{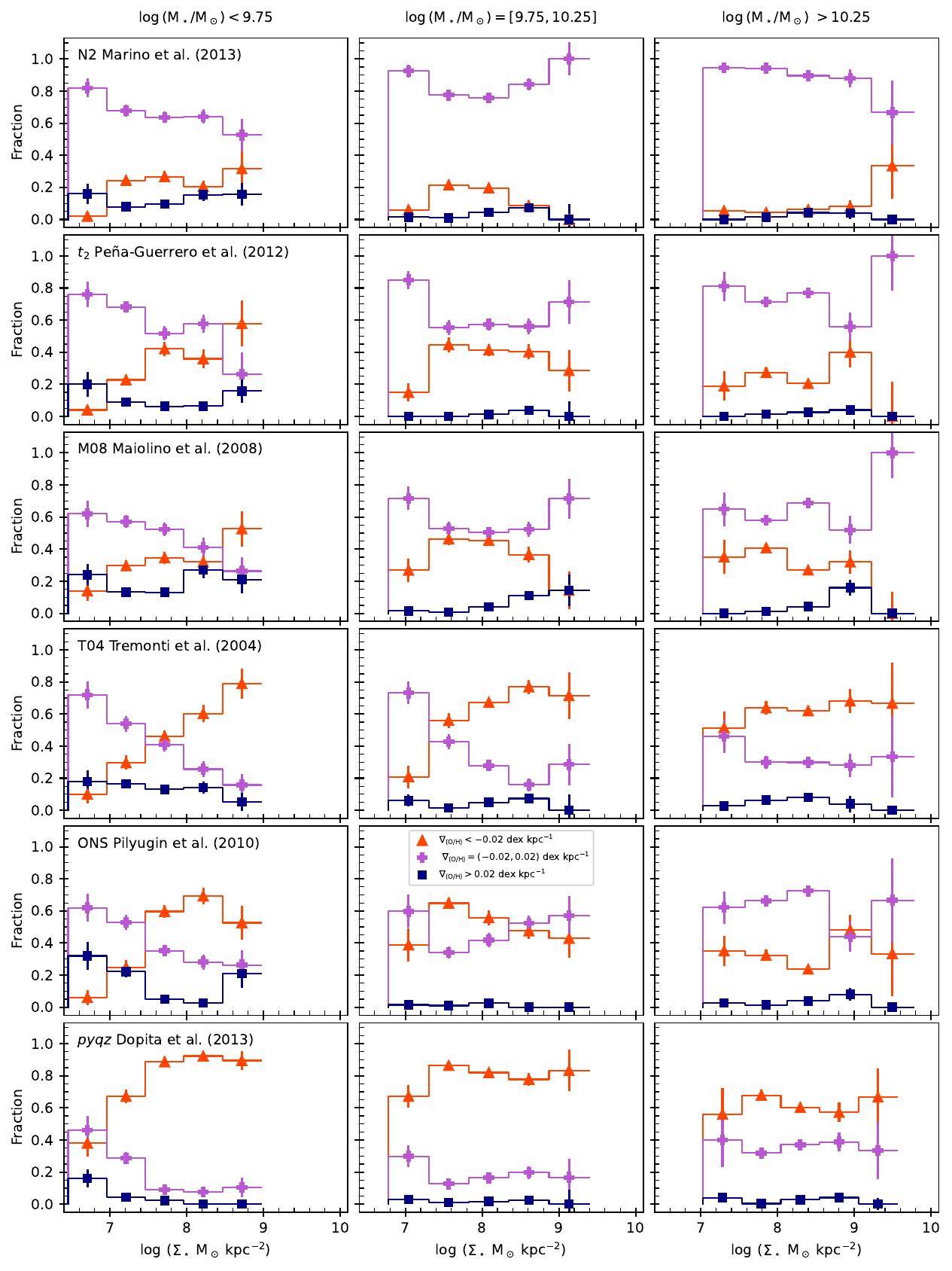}}
%\resizebox{8cm}{!}{\includegraphics{Figuras/pdf/l25z+-vsig-sigma.pdf}}
%\resizebox{8.cm}{!}{\includegraphics{Figuras/pdf/l25z0-mzr-grads_0.02.pdf}}
\caption{Frequency of galaxies with different metallicity gradients as a function of \sigmastar~for our MaNGA galaxies. Comparison of the trends generated by adopting six additional metallicity calibrators. Orange, purple and blue shaded lines represent the fraction of galaxies with strong negative, weak and strong positive metallicity gradients,
respectively. The sample was subdivided in low- (left panels), intermediate- (centrals panels) and high- (right panels) mass galaxies. All error bars were estimated by using a  bootstrap technique.}
\label{frequencycalibrators}
\end{figure*}

As can be seen from Fig.~\ref{mzrdifcalibrators}, the results do not depend strongly on the chosen metallicity calibrator except for N2 \citep{marino2013} which yields no trend between sSFR and \sigmastar~(Fig.~\ref{sigmastarcalibrators}). We note that,  as was shown by \citet{zhang2017}, the abundances measured using N2 calibrator can be affected by DIG (Diffused Ionised Gas) contribution. This might explain the different behaviours obtained in this case.  On the other hand, the frequencies of galaxies in each gradient interval exhibit larger variations as a function of \sigmastar~(Fig.~\ref{frequencycalibrators}).  Certainly, the metallicity calibrators remain an important issue for the determination of the metallicity gradients. 

%\begin{figure}
%\resizebox{8cm}{!}{\includegraphics{Figuras/pdf/mzr-calibrators.pdf}}
%\caption{Direct comparison of the shape of the MZR built from seven different metallicity calibrators available in MaNGA database. The %calibrator used in this work is shown in grey stars.}
%\label{}
%\end{figure}

%\begin{figure}
%\resizebox{8cm}{!}{\includegraphics{Figuras/pdf/manga-vsig-sigma-colored.pdf}}
%----------
%\resizebox{8cm}{!}{\includegraphics{Figuras/pdf/manga-vsig-sigma.pdf}}
%\resizebox{8.cm}{!}{\includegraphics{Figuras/pdf/l25z0-mzr-grads_0.02.pdf}}
%--------------
%\caption{ , respectively. All error bars were estimated by using a  bootstrap technique.}
%\label{mangavsig}
%\end{figure}

\end{document}